\documentclass[draftcls, 12pt,onecolumn,oneside,compress]{IEEEtran}
\usepackage{geometry}
\pdfoutput=1

\ifCLASSINFOpdf
\else
   \usepackage[dvips]{graphicx}
\fi
\usepackage{url}

\usepackage{caption}
\usepackage{amssymb}
\usepackage{graphicx,color}
\usepackage{mathrsfs}
\usepackage{bm}
\usepackage{amsmath}
\usepackage{amsfonts,amssymb}
\usepackage{extarrows}
\usepackage{graphicx}
\usepackage{algorithm}
\usepackage{algorithmic}
\usepackage{setspace}
\usepackage{array}
\usepackage{epstopdf}
\usepackage{subfigure}
\usepackage{caption}
\usepackage{float}

\newtheorem{lemma}{Lemma}
\newenvironment{proof}{{\it Proof\,:}\;}{\hfill $\square$\par}
\newtheorem{assumption}{Assumption}
\usepackage{cite}
\usepackage{url}
\usepackage{stfloats}
\DeclareMathOperator*{\argmax}{arg\,max}

\begin{document}

\title{Deep Learning based Antenna Selection and CSI Extrapolation in Massive MIMO Systems}
\author{Bo Lin, \IEEEmembership{Student Member, IEEE}, Feifei Gao, \IEEEmembership{Fellow, IEEE}, Shun Zhang, \IEEEmembership{Senior Member, IEEE}, Ting Zhou, and Ahmed Alkhateeb
\thanks{
B. Lin and F. Gao are with Department of Automation, Tsinghua University, State Key Lab of Intelligent Technologies and Systems, Tsinghua University, State Key for Information Science and Technology (TNList) Beijing 100084, P. R. China (e-mail: feifeigao@ieee.org; linb20@mails.tsinghua.edu.cn).
S. Zhang is with the State Key Laboratory of Integrated Services Networks, Xidian University, Xian 710071, P.R. China (e-mail: zhangshunsdu@xidian.edu.cn).
T. Zhou is with the Shanghai Frontier Innovation Research Institute, Chinese Academy of Sciences, Shanghai 201210, P.R. China (e-mail: zhouting@sari.ac.cn).
A. Alkhateeb is with the School of Electrical, Computer and Energy Engineering, Arizona State University, Tempe, AZ 85287 USA (e-mail: alkhateeb@asu.edu). }\vspace{-15mm}}
\maketitle
\thispagestyle{empty}
\begin{abstract}
A critical bottleneck of massive multiple-input multiple-output (MIMO) system is the huge training overhead caused by downlink transmission, like channel estimation, downlink beamforming and covariance observation.
In this paper, we propose to use the channel state information (CSI) of a small number of antennas to extrapolate the CSI of the other antennas and reduce the training overhead. Specifically, we design a deep neural network that we call an antenna domain extrapolation network (ADEN) that can exploit the correlation function among antennas.
We then propose a deep learning (DL) based antenna selection network (ASN) that can select a limited antennas for optimizing the extrapolation, which is conventionally a type of combinatorial optimization and is difficult to solve. We trickly designed a constrained degradation algorithm to generate a differentiable approximation of the discrete antenna selection vector such that the back-propagation of the neural network can be guaranteed.
Numerical results show that the proposed ADEN  outperforms the traditional fully connected one, and the antenna selection scheme learned by ASN is much better than the trivially used uniform selection.
\end{abstract}

\begin{IEEEkeywords}
Channel extrapolation,  deep learning, antenna selection,  channel covariance matrix, beam prediction
\end{IEEEkeywords}

\IEEEpeerreviewmaketitle

\section{Introduction}\label{section1}
Massive MIMO has attracted tremendous attention in the area of wireless communications, in which the base station (BS) is equipped with a large scale of antennas and can simultaneously serve multiple users.
It is well admitted that massive MIMO could significantly boost the system capacity and transmission rate, making it a promising technique for both 5G and future  wireless communications \cite{larsson2014massive}.
Nevertheless, accurate downlink channel state information (CSI) is the prerequisite for achieving the full potential of massive MIMO, and the pilot length is proportional to  the number of transmit antennas.
Hence, the training overhead for downlink transmission   becomes extremely large.

Most existing works assumed sparsity when performing the channel estimation, since the BS is always deployed in a high place and the massive MIMO system mostly  works in millimeter wave (mmWave) frequency band \cite{xie2016unified}.
In turn, many channel estimation algorithms such as compressive sensing (CS) methods \cite{bajwa2010compressed,steffens2016multidimensional,cheng2013channel} and angle domain MIMO channel reconstruction \cite{wang2018spatial,jian2019angle,zhao2017angle} have been explored.
However, these approaches mainly rely on simple mathematical models, which may not be accurate in complicated channel environment.

Recently, deep learning (DL), a new artificial intelligence (AI) method, has demonstrated its powerful advantages in many research areas, like image processing, speech processing, and natural language processing.
The application of DL in physical layer communications is also sweeping \cite{qin2019deep}, and many efforts have been made in channel estimation\cite{he2018deep,ye2017power}, signal detection\cite{ye2017power,he2019model}, and beam prediction \cite{zhou2018deep,klautau20185g}, etc.
In terms of saving the training overhead, a number of DL based channel prediction methods have been proposed \cite{guo2020convolutional,ma2020sparse} and achieved better results than the traditional methods.
In \cite{dong2018machine}, Dong et al. designed  a machine learning method to predict the channel of a part of antennas from that of the other antennas, where channel prediction is modeled as a linear function and is solved by linear regression (LR) and support vector regression (SVR).
In \cite{alrabeiah2019deep}, Alrabeiah  et al. raised the concept of channel extrapolation  in space and frequency.
An important observation made in \cite{alrabeiah2019deep} is that there exists an implicit mapping function between the channels of two antenna sets with different frequencies and positions as long as the position-to-channel mapping is bijective.
Subsequently, Taha et al. proposed a DL based method to find the optimal reconfigurable intelligent surface (RIS) reflection matrices that approaches the maximum achievable rate with only a few  active antenna elements \cite{taha2019deep}. Moreover,
Yang et al. predicted the downlink channel from the uplink channel for FDD massive MIMO systems with acceptable accuracy \cite{yang2019deep}.

Nevertheless, \cite{alrabeiah2019deep,taha2019deep,yang2019deep,yang2019deeptransfer} simply applied the fully-connected  deep neural networks (DNN), while few specific structure has been designed to fit the antenna domain channel extrapolation.
Moreover, it is readily known that different antenna selection scheme will achieve different extrapolation accuracy, while the existing works \cite{dong2018machine,alrabeiah2019deep,taha2019deep}  simply  adopted the uniformly selected antennas for channel extrapolation.
Although uniform selection is effective in many traditional works, it cannot guarantee the optimality due to the various electromagnetic field characteristics of the environment.
It is also noted that many traditional antenna selection algorithms \cite{molisch2004mimo,sanayei2004antenna,molisch2005capacity} refer to selecting the optimal antennas for data transmission after the channels are known, which is obviously not applicable for channel extrapolation.

In this paper, we propose an antenna domain extrapolation network (ADEN) to perform channel extrapolation and an antenna selection network (ASN) to choose the optimal antennas for the extrapolation. Specifically, the antenna domain extrapolation is divided into two steps, that is coarse CSI extrapolation and fine CSI extrapolation. The coarse CSI extrapolation is realized by a fully connected neural network and the fine CSI extrapolation is modeled as an ordinary differential equation (ODE) initial value problem, where the coarse extrapolated CSI is the initial value while the fine extrapolated CSI is the final value. Moreover, a key challenge of ASN is that the operation of selecting antennas is non-differentiable and cannot guarantee the back-propagation. We then design a constrained degradation algorithm (CDA)  that formulates a differentiable approximation of the antenna selection vector. To enhance the overall performance, we next propose to train the ASN and ADEN jointly by penalizing both the extrapolation error and the antenna selection vector. Based on ASN and ADEN, we present three typical applications: (i) extrapolating CSI from a part of antennas to all antennas; (ii) extrapolating channel covariance matrix (CCM) from a part of antennas to all antennas; (iii) using CSI from a part of antennas to predict the downlink beamforming coefficient of all antennas directly.
The simulation results show that the proposed ADEN is better than the existing fully connected DNN, and the proposed ASN is much better than trivially accepted uniform selection.

The remainder of this paper is organized as follows.
Section \ref{section2} introduces the system and channel model.
Section \ref{section3} designs the ASN and the ADEN.
Section \ref{section4} presents  three typical applications of the CSI extrapolation.
Section \ref{section5}  provides the simulation results and   Section \ref{section6} draws the conclusion.

\textbf{Notation}:
Bold uppercase $\mathbf{A}$ is a matrix, bold lowercase $\bm{a}$ is a column vector, non-bold letter $a$ and $A$ are scalars, caligraphic letter $\mathcal{A}$ is a set;
$|a|$ is a magnitude of a scalar, $\|\mathbf{a}\|_{p}$ is the p-norm of a vector, $\|\mathbf{A}\|_{F}$ is the Frobenius norm of a matrix, $|\mathcal{A}|$ is the cardinality of a set;
$\mathbf{A}^{T}$, $\mathbf{A}^{*}$, $\mathbf{A}^{H}$ are the Hermitian, conjugate, and transpose of $\mathbf{A}$;
$\odot$ and $\otimes$ represent the Hadamard product and Kronecker product operator respectively;
$\Re(\mathbf{s})$ and $\Im(\mathbf{s})$ are the real and imaginary component of $\mathbf{s}$;
$\mathbb{E}$ is the expectation.
\section{System and Channel Model}\label{section2}

\subsection{System Model}
Let us consider a system where a BS is communicating with a mobile user.
The BS has $N_t\gg 1$ antennas and the user has only one antenna.
Denote $\mathbf{h}\in \mathbb{C}^{N_{t}}$ as the downlink channel vector from the BS  to the user,
$\mathbf{P}\in \mathbb{C}^{N\times N_{t}}$ as the downlink pilot   matrix with $N$ being the length of pilots, and $\mathbf{v}\in \mathbb{C}^{N}$ as the vector of sensor noise with power $\sigma^2$.
The received signal at the user is
\begin{equation}\label{receive_dignal}
\mathbf{y} = \mathbf{P}\mathbf{h} + \mathbf{v}.
\end{equation}

There are many traditional methods to perform channel estimation such as least square (LS) channel estimation and linear minimum mean square error (LMMSE) channel estimation.
The LS channel estimation can be formulated as
\begin{equation}\label{LS}
\mathbf{h}_{LS} = \mathbf{P}^{\dagger}\mathbf{y},
\end{equation}
where $\mathbf{P}^{\dagger}=\mathbf{P}^{H}\left(\mathbf{P}\mathbf{P}^{H}\right)^{-1}$ is the pseudo-inverse of the matrix $\mathbf{P}$.
When the signal-to-noise ratio (SNR) is not high enough, LS estimation will bring a large estimation error.
In this case, LMMSE channel estimation could be adopted to obtain higher estimation accuracy:
\begin{equation}\label{LMMSE}
\mathbf{h}_{MMSE} = \left(\mathbf{y}^{T}(\mathbf{P}^{H}\mathbf{R}_{\rm h}\mathbf{P}+\sigma^{2}\mathbf{I})^{-1}\mathbf{P}^{H}\mathbf{R}_{\rm h}\right)^{T},
\end{equation}
where  $\mathbf{I}\in\mathbb{C}^{N\times N}$ is the identity matrix.
Nevertheless, to perform LMMSE channel estimation, the statistical CSI, i.e.,  CCM $\mathbf{R}_{\rm h} =\mathbb{E}\left[\mathbf{h}\cdot \mathbf{h}^{H}\right]\in \mathbb{C}^{N_{t}\times N_{t}}$ is needed.
From (\ref{LS}) and (\ref{LMMSE}), we see that the training  consumption is extremely high for massive number of antennas.
A natural question then arises:
Can we use the channel of $M_{t} (M_{t}\textless N_{t})$ BS antennas to recover the channel of all $N_{t}$ antennas?

\textbf{Extrapolation Based Channel prediction:}
Define $\mathcal{A}\in \mathbb{Z}^{N_{t}\times1}$ as the complete set of all antennas and $\mathcal{B}\in \mathbb{Z}^{M_{t}\times1}$ as a subset of $\mathcal{A}$ with size $M_t$.
Moreover, denote $\mathbf{h}_{\mathcal{A}}$ (same as $\mathbf{h}$ in (\ref{receive_dignal})) as the vector that contains the channel of antenna set $\mathcal{A}$ and $\mathbf{h}_{\mathcal{B}}$ as the subset of $\mathbf{h}_{\mathcal{A}}$ that contains the channel of antenna set $\mathcal{B}$.
It has been proved in \cite{alrabeiah2019deep} that if the position-to-channel mapping is bijective, then the channel-to-channel mapping exists.
For a given static communication environment including the geometry, materials, antenna positions, etc., the location of the user and the channel usually correspond strictly, i.e., the position-to-channel mapping function is usually bijective \cite{alrabeiah2019deep}.
Hence the following channel mapping exists
\begin{equation}\label{channel-mapping}
\bm{\Phi}_{\mathbf{h}}:\left\{\mathbf{h}_{\mathcal{B}}\right\} \rightarrow \left\{\mathbf{h}_{\mathcal{A}}\right\}.
\end{equation}

\textbf{Extrapolation Based Beam Prediction:}
In massive MIMO systems, downlink beamforming is necessary for spatial multiplexing.
The optimal beam for channel $\mathbf{h}_{\mathcal{A}}$ is chosen from the beamforming codebook $\mathbf{F}=\left\{\mathbf{f}_{1},\mathbf{f}_{2},\cdots,\mathbf{f}_{|\mathbf{F}|}\right\}$  that maximizes the system rate
\begin{equation}\label{beam}
\mathbf{f}_{\mathcal{A}}=\argmax_{\mathbf{f} \in {\mathbf{F}}}\text{log}_{2}\left(1+\text{SNR}\left|\mathbf{h}_{\mathcal{A}}^{T}\mathbf{f}\right|^{2}\right).
\end{equation}
The number of beamforming vectors in the codebook is proportional to the number of antennas at BS.
Hence, the time and computation cost when selecting the optimal beam is also large in massive MIMO system.
To save time and computation resources, we propose the beam extrapolation that predict the downlink beamforming of all antennas from a part of antennas' channel,
which utilizes the channel of a part of antennas to predict the beam index of the whole antennas.
In fact, from (\ref{channel-mapping}) and (\ref{beam}), we know that the channel-to-beam mapping exists and can be denoted as
\begin{equation}\label{channel-beam}
\bm{\Phi}_{\mathbf{beam}}:\left\{\mathbf{h}_{\mathcal{B}}\right\} \rightarrow \left\{\mathbf{f}_{\mathcal{A}}\right\}.
\end{equation}

\textbf{Extrapolation Based Covariance Prediction:}
In addition to channel and beam, CCM is also an important parameter for transceiver design.
For example, CCM is used both to design optimal pilots and compute the LMMSE channel estimation.
CCM can also be used to find the subspace of the beamforming vector in a coarse and blind way.
However, the cost of obtaining CCM is huge.
We then propose to utilize the CCM of a part of antennas to extrapolate the CCM of the whole antennas.
Before proving the existence of the CCM-to-CCM mapping, we adopt the following assumption:

\begin{assumption}
The mapping $\bm{g}_{\mathcal{B}}: \left\{\mathcal{C}\right\} \rightarrow \left\{\mathbf{R}_{\mathcal{B}}\right\}$ is bijective,
where $\mathcal{C}$ denotes the location of an area, and $\mathbf{R}_{\mathcal{B}}$ denotes the CCM of the area and the antenna set $\mathcal{B}$.
\end{assumption}

Assumption 1 means that each area in the candidate set $\left\{\mathcal{C}\right\}$ has a unique CCM.
Note that the bijectiveness of mapping $\bm{g}_{\mathcal{B}}$ depends on some truths including: (i) the signal attenuation from the BS to different areas is different; (ii) the  geometry and materials of different areas are different; (iii) the scattering paths in different areas are different.

Now, the inverse mapping of $\bm{g}_{\mathcal{B}}$ can be described as
\begin{equation}\label{g-ccm}
\bm{g}_{\mathcal{B}}^{-1}:\left\{\mathbf{R}_{\mathcal{B}}\right\} \rightarrow \left\{\mathcal{C}\right\}.
\end{equation}

For antenna set $\mathcal{A}$, there also exists a mapping $\bm{g}_{\mathcal{A}}: \left\{\mathcal{C}\right\} \rightarrow \left\{\mathbf{R}_{\mathcal{A}}\right\}$. Hence, the CCM-to-CCM mapping exist, i.e.,
\begin{equation}\label{mapping-ccm}
\bm{\Phi}_{\mathbf{R}}=\bm{g}_{\mathcal{A}}\circ \bm{g}_{\mathcal{B}}^{-1}=\left\{\mathbf{R}_{\mathcal{B}}\right\} \rightarrow \left\{\mathbf{R}_{\mathcal{A}}\right\}.
\end{equation}

The previously described channel-to-channel mapping (\ref{channel-mapping}), channel-to-beam mapping (\ref{channel-beam}), and CCM-to-CCM mapping (\ref{mapping-ccm}) can be summarized in a unified extrapolation function
\begin{equation}\label{mapping}
\bm{\Phi}:\left\{\mathbf{u}_{\mathcal{B}}\right\} \rightarrow \left\{\mathbf{u}_{\mathcal{A}}\right\},
\end{equation}
where $\mathbf{u}_{\mathcal{B}}$ denotes the information of antenna set $\mathcal{B}$ and $\mathbf{u}_{\mathcal{A}}$ denotes the information of antenna set $\mathcal{A}$.
As the exact mathematical function of extrapolation is hardly to obtain\footnote{The mapping can be treated as the interpolation in antenna domain. The interpolation operation is always based on an explicit model. For example, The interpolation on the orthogonal frequency division multiplexing (OFDM) subcarrier is realized by using discrete Fourier transform (DFT). However, in the antenna domain, there is no explicit model to describe the mapping (\ref{mapping}). Hence, the linear interpolation result will be very poor.},
we adopt deep neural networks (DNN) to fit such function with the aided of training data.
Then, the extrapolation function can be described as
\begin{equation}\label{extrapolation_function}
\left\{\mathbf{u}_{\mathcal{A}}\right\} = f(\left\{\mathbf{u}_{\mathcal{B}}\right\},\bm{\Theta_{e}}),
\end{equation}
where $\bm{\Theta_{e}}$ is the parameters of DNN.

\begin{figure}[t]
\centering
\includegraphics[width=0.8\textwidth]{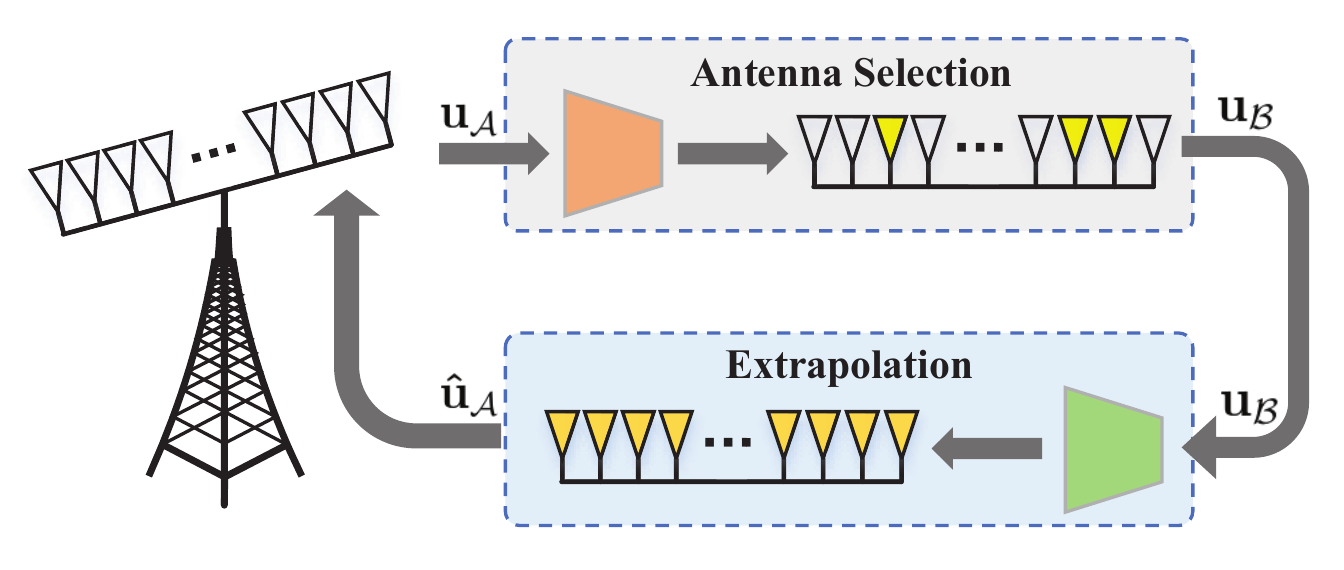}
\caption{Block diagram of the described antenna selection and extrapolation model.}
\label{block_diagram}
\end{figure}

\subsection{Channel Model}
We adopt a 3-D geometric based channel model \cite{samimi20163} where signal emitted by the transmitter reaches the receiver from multiple paths through reflection, diffraction, and refraction \cite{sayeed2010wireless}.
Denote $\alpha_{l}$ as the attenuation coefficient of the $l$-th path, $\phi_{l}^{a,D}$ as the azimuth angle of departure (AoD) of the $l$-th path, $\phi_{l}^{e,D}$ as the elevation AoD for the $l$-th path, $\phi_{l}^{a,A}$ as the azimuth angle of arrival (AoA) of the $l$-th path, $\phi_{l}^{e,A}$ as the elevation AoA of the $l$-th path, $\vartheta_{l}$ as the phase of path $l$ and $\tau_{l}$ as the propagation delay of the $l$-th path.
The channel vector $\mathbf{h}$ is given by \cite{sayeed2002deconstructing}
\begin{equation}\label{channelmodel}
\mathbf{h} = \sum_{l=1}^{L}\alpha_{l}e^{j(\vartheta_{l}+2\pi \tau_{l}B)}\mathbf{a}(\phi_{l}^{a,A},\phi_{l}^{e,A})\mathbf{a}^{*}(\phi_{l}^{a,D},\phi_{l}^{e,D}),
\end{equation}
where $B$ is the signal bandwidth, and $\mathbf{a}(\phi_{l}^{a,A},\phi_{l}^{e,A})$ and $\mathbf{a}(\phi_{l}^{a,D},\phi_{l}^{e,D})$ are the steering vectors at the arrival and departure sides.
The mathematical expression of $\mathbf{a}(\cdot)$ is
\begin{equation}
\mathbf{a}(\phi_{l}^{a,A},\phi_{l}^{e,A}) = \mathbf{a}_{z}(\phi_{l}^{e,A})\otimes \mathbf{a}_{y}(\phi_{l}^{a,A},\phi_{l}^{e,A})\otimes \mathbf{a}_{x}(\phi_{l}^{a,A},\phi_{l}^{e,A}),
\end{equation}
where $\mathbf{a}_{x}(\cdot)$, $\mathbf{a}_{y}(\cdot)$, $\mathbf{a}_{z}(\cdot)$ are the BS array response vectors in the $x$, $y$, and $z$ directions (the operation is the same for the AoD). Moreover, the operators $\mathbf{a}_{x}(\cdot)$, $\mathbf{a}_{y}(\cdot)$, $\mathbf{a}_{z}(\cdot)$ are respectively defined as
\begin{equation}
\begin{aligned}
\mathbf{a}_{x}(\phi_{l}^{a,A},\phi_{l}^{e,A}) &= [1,e^{j\frac{d_{x}}{\lambda}sin(\phi_{l}^{e,A})cos(\phi_{l}^{a,A})},\cdots,e^{j\frac{d_{x}}{\lambda}(N_{x}-1)sin(\phi_{l}^{e,A})cos(\phi_{l}^{a,A})}],\\
\mathbf{a}_{y}(\phi_{l}^{a,A},\phi_{l}^{e,A}) &= [1,e^{j\frac{d_{y}}{\lambda}sin(\phi_{l}^{e,A})sin(\phi_{l}^{a,A})},\cdots,e^{j\frac{d_{y}}{\lambda}(N_{y}-1)sin(\phi_{l}^{e,A})sin(\phi_{l}^{a,A})}],\\
\mathbf{a}_{z}(\phi_{l}^{e,A}) &= [1,e^{j\frac{d_{z}}{\lambda}cos(\phi_{l}^{e,A})},\cdots,e^{j\frac{d_{z}}{\lambda}(N_{z}-1)cos(\phi_{l}^{e,A})}],
\end{aligned}
\end{equation}
where $\lambda$ is the carrier wavelength, while $d_{x}$, $d_{y}$, $d_{z}$ are the antenna spacings in the $x$-, $y$-, and $z$- direction.

\section{Deep Learning Based Antenna Selection and Antenna Domain Extrapolation}\label{section3}
\begin{figure}[t]
\centering 
\includegraphics[width=0.8\textwidth]{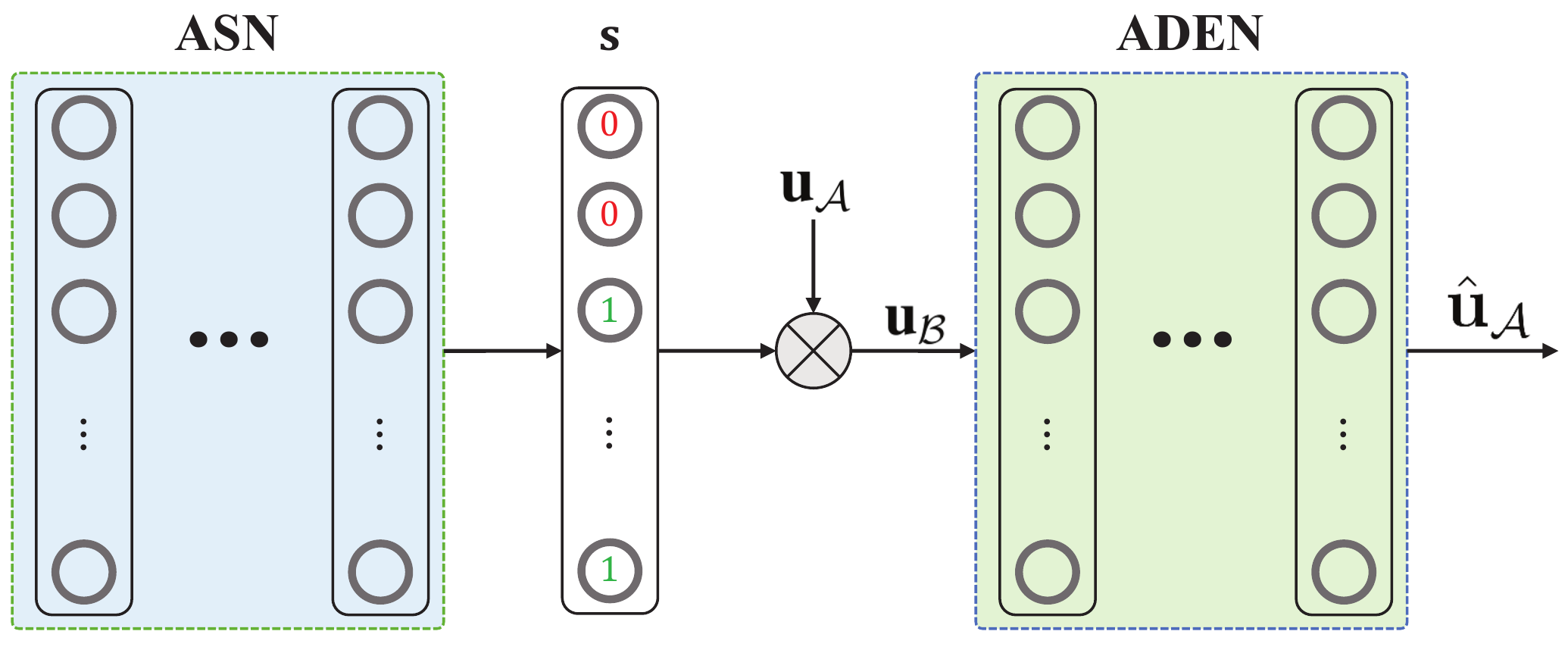}
\caption{The overall structure of ASN and ADEN.}
\label{jointasnaen}   
\end{figure}
We here propose a DL based joint design that contains two subnetworks,
antenna selection network (ASN) and antenna domain extrapolation network (ADEN),  as shown in Fig. \ref{jointasnaen}.
Define the output of ASN as the antenna selection vector $\textbf{s}=\left\{s_{1},s_{2},\cdots s_{N_{t}}\right\}\in \left\{0,1\right\}^{N_{t}}$ that is an $M_{t}$-hot vector with $M_{t}$ elements being `$1$' and the other elements being `$0$'.
Specifically, we set $s_{i}={1}$ if the $i\text{-th}$ antenna is selected, while $s_{i}={0}$ otherwise.
The input of ADEN is $\mathbf{u}_{\mathcal{B}}=\mathbf{s}\odot\mathbf{u}_{\mathcal{A}}$, and the output of ADEN is the extrapolated information  $\hat{\mathbf{u}}_{\mathcal{A}}$.
The ASN is trained to find the antenna selection vector that minimizes the extrapolation error of the ADEN.
We also propose to connect the ASN and ADEN through a product operation, and then jointly train them via backpropagation at the same time.

\subsection{Antenna Selection Network}
\label{sec:guidelines}

\begin{figure*}
\centering
\includegraphics[width=0.8\textwidth]{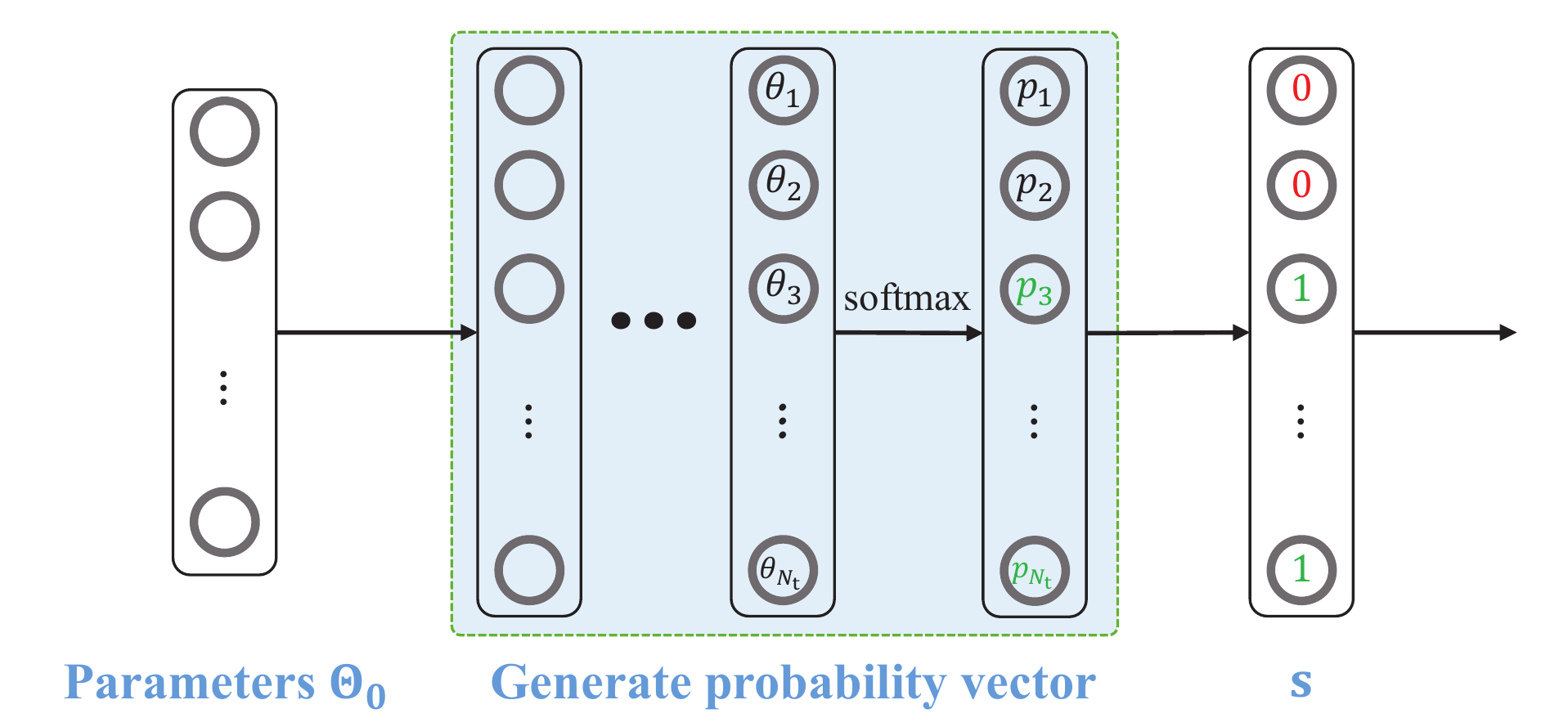}
\caption{The three parts of ASN.}
\label{ASN}
\end{figure*}

The ASN is composed of three parts as shown in Fig. \ref{ASN}.
The first part is a layer of randomly initialized parameters $\bm\theta_{\bm 0}$.
The second part contains several layers of fully connected neurons to generate a probability vector $\mathbf{p}=\left\{p_{1},p_{2},\cdots,p_{N_{t}}\right\}$, where $p_{i}$ represents the probability of the $i$-th antenna being selected.
The overall vector $\mathbf{p}$ satisfies the condition $\sum_{i=1}^{N_{t}}p_{i}=1$.
Denote the output of the layer before the probability layer (also the input of probability vector) as $\bm\theta_{\mathbf{p}}=\left\{\theta_{1},\theta_{2},\cdots,\theta_{N_{t}}\right\}$.
Then $p_{i}$ is generated by
\begin{equation}\label{softmax}
\begin{aligned}
p_{i}=\text{softmax}\left(\bm\theta_{\mathbf{p}}\right)_{i}=\frac{\text{exp}\left(\theta_{i}\right)}{\sum\limits_{j=1}^{N_{t}}\text{exp}\left(\theta_{j}\right)}.
\end{aligned}
\end{equation}

The third part is the antenna selection vector $\textbf{s}$ that is generated based on $\mathbf{p}$.
Specially, define the index of the biggest $M_{t}$ elements as
\begin{equation}\label{index}
\mathcal{I}_{\mathbf{s}} = \mathop{\rm arg\,top\,M_{t}}\left\{\mathbf{p}\right\},
\end{equation}
where $\mathop{\rm arg\,top\,M_{t}}\left\{\mathbf{x}\right\}$ is a function that finds the biggest $M_{t}$ elements in vector $\mathbf{x}$, and $\mathcal{I}_{\mathbf{s}}$ is an index set.
The elements of $\mathbf{s}$ with index $\mathcal{I}_{\mathbf{s}}$ are $1$ and otherwise are $0$.

We adopt back-propagation algorithm to train the ASN that requires all operations in the neural network being
differentiable.
However, when generating $\mathbf{s}$, the function $\mathop{\rm arg\,top\,M_{t}}\left\{\cdot\right\}$ is not differentiable, which is the key obstacle of performing the antenna selection via DL.
To solve this problem, let us first provide the following lemma:
\begin{lemma}\label{lemma1}
For two positive integers $M_{t}$ and $N_{t}$ with $
M_{t}\textless N_{t}$,  the vector $\bm{\nu}=\{[\nu_{1},\cdots,\nu_{N_{t}}]^{T}|\nu_{1}+\nu_{2}+\cdots+\nu_{N_{t}}=M_{t}, \forall i,\nu_{i}\in \mathbb{R},0\leq\nu_{i}\leq M_{t}\}$ satisfies the following equality constraints
\begin{equation}\label{constraint}
\begin{aligned}
\left\{
\begin{aligned}
\nu_{1}^{2}+\nu_{2}^{2}+\cdots+\nu_{N_{t}}^{2}=M_{t}\\
\nu_{1}^{3}+\nu_{2}^{3}+\cdots+\nu_{N_{t}}^{3}=M_{t}\\
\end{aligned}
\right..
\end{aligned}
\end{equation}
Let us sort the elements of vector $\bm\nu$ in descending order as $\bm\nu^{'}=\left\{\nu_{k_{1}},\nu_{k_{2}},\cdots,\nu_{k_{N_{t}}}\right\}$, where $\nu_{k_{1}}\geq\nu_{k_{2}}\geq\cdots\nu_{k_{N_{t}}}\geq0$.
Then $\bm\nu$ is an $M_{t}$-hot vector and $\nu_{i}$ has the following form
\begin{equation}\label{nu}
\begin{aligned}
\nu_{i}=&\left\{
\begin{aligned}
&1,\; &\,&i=k_{1},k_{2},\cdots,k_{M_{t}}\\
&0,\; &\,&i=k_{M_{t}+1},k_{M_{t}+2},\cdots,k_{N_{t}}
\end{aligned}
\right..
\end{aligned}
\end{equation}
\end{lemma}

\begin{figure*}[t]
\centering
\includegraphics[width=0.85\textwidth]{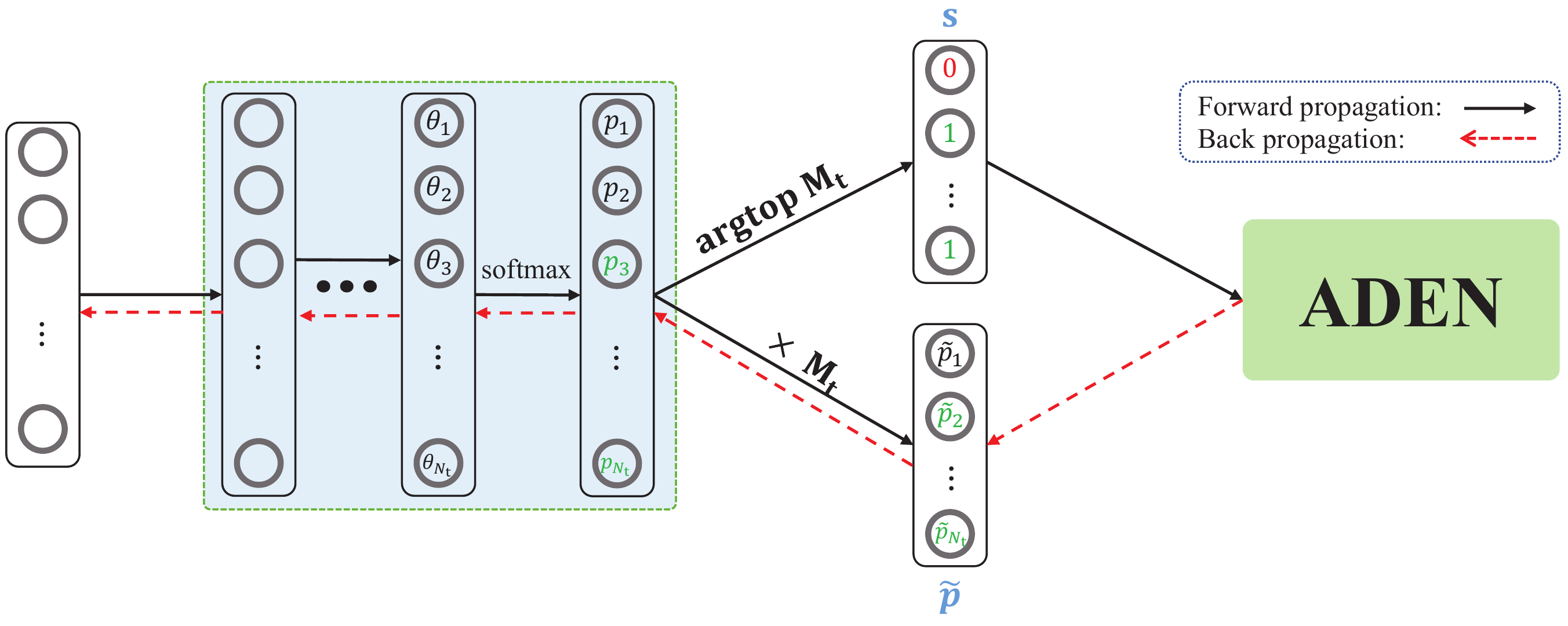}
\caption{Constrained degradation algorithm: forward-propagation link and back-propagation link of ASN.}
\label{backpropagation}
\end{figure*}

The proof of Lemma \ref{lemma1} can be referred to in Appendix \ref{appendix_proof}, and the geometric interpretation (or inspiration) of the Lemma 1 is shown in Appendix \ref{appendix_lemma}.
Based on Lemma \ref{lemma1}, we design a  constrained degradation algorithm (CDA)  that could construct a differentiable approximation of the non-differentiable vector $\mathbf{s}$.

According to the definition (\ref{softmax}), the softmax output $\mathbf{p}$ satisfies $\sum_{i=1}^{N_{t}}p_{i}=1$ and $0\leq p_{i}\leq 1$.
Let us construct $\widetilde{\mathbf{p}}=\left\{\widetilde{p_{i}}=M_{t}\cdot p_{i}|i=1,\cdots,N_{t}\right\}$.
Obviously $\widetilde{\mathbf{p}}$ satisfies $\sum_{i=1}^{N_{t}}\widetilde{p}_{i}=M_{t}$ and $0\leq \widetilde{p_{i}}\leq M_{t}$.
From Lemma \ref{lemma1}, we know that if $\widetilde{\mathbf{p}}$ satisfies the equality constraints (\ref{constraint}), then $\widetilde{\mathbf{p}}$ is an $M_{t}$-hot vector.
However, the deterministic constraints (\ref{constraint}) are difficult to implement in neural networks.
We then adopt the following penalty to make $\widetilde{\mathbf{p}}$ gradually  satisfy constraints (\ref{constraint}):
\begin{equation}\label{penalty_asn}
\mathcal{L}_{ASN}=\alpha_{1}\left(\|\widetilde{\mathbf{p}}\|_{2}^{2}-M_{t}\right)^{2}+\alpha_{2}\left(\|\widetilde{\mathbf{p}}\|_{3}^{3}-M_{t}\right)^{2},
\end{equation}
where $\alpha_{1}\textgreater0$ and $\alpha_{2}\textgreater0$ are the tuning parameters of the two parts of penalties.
During the training process, the penalty $\mathcal{L}_{ASN}$ keeps on decreasing and will approach zero.
Hence, both $\|\widetilde{\mathbf{p}}\|_{2}^{2}-M_{t}$ and $\|\widetilde{\mathbf{p}}\|_{3}^{3}-M_{t}$ will approach zero, and $\widetilde{\mathbf{p}}$ will tend to satisfy constraints (\ref{constraint}), i.e., $\widetilde{\mathbf{p}}$ will tend to be $\mathbf{s}$.
Interestingly, $\widetilde{\mathbf{p}}$ will be always differentiable when it gradually approaches $\mathbf{s}$.
Hence, the key idea of CDA is to utilize the differentiable $\widetilde{\mathbf{p}}$ as an approximation of the non-differentiable $\mathbf{s}$ during back-propagation.
The forward-propagation and back-propagation links are summarized in Fig. \ref{backpropagation}.

\subsection{Antenna Domain Extrapolation Network}
\begin{figure*}[t]
\centering
\includegraphics[width=0.9\textwidth]{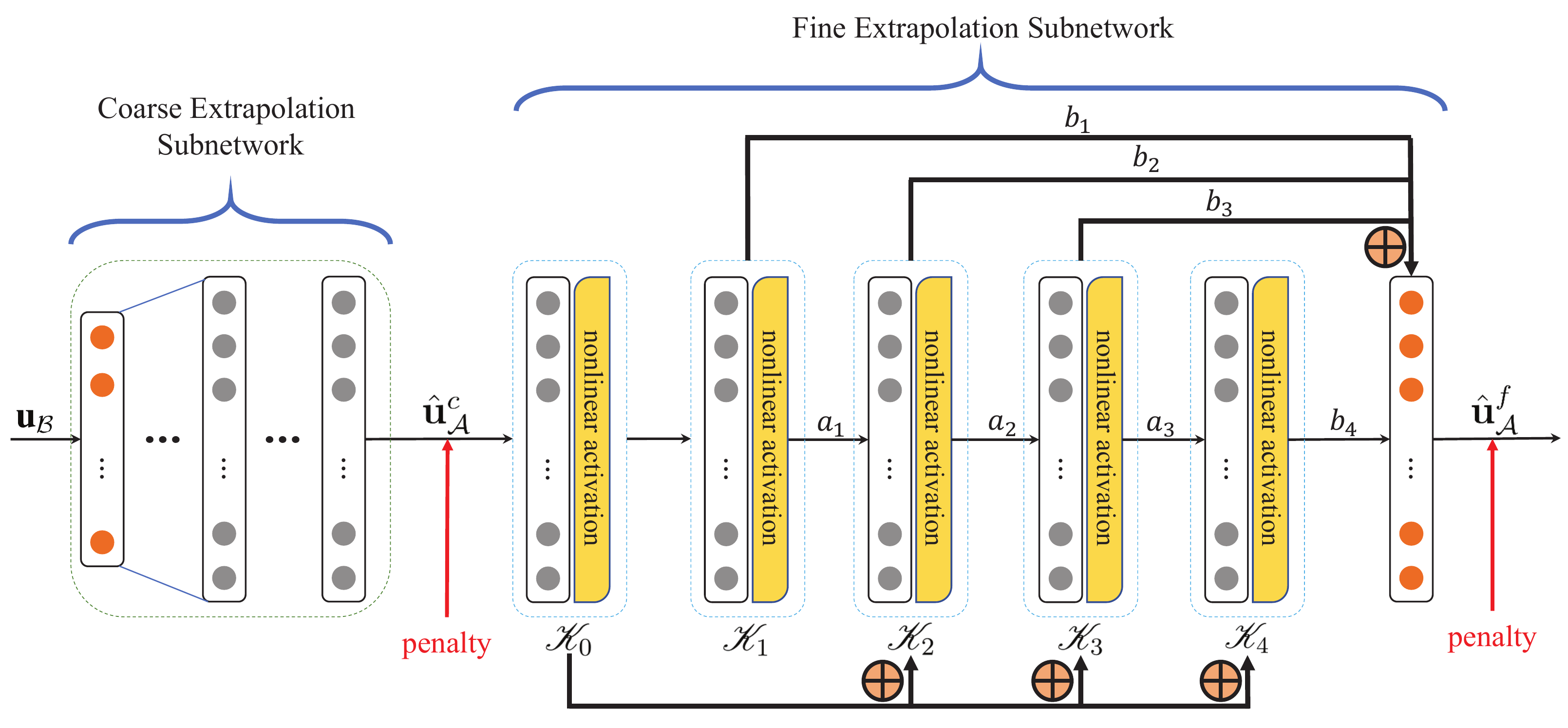}
\caption{Antenna domain extrapolation network.}
\label{ADEN}
\end{figure*}
The ADEN is composed of two parts as shown in Fig. \ref{ADEN}.
The first part is coarse extrapolation subnetwork that incudes several fully connected neural layers.
The second part is fine extrapolation subnetwork that improves the extrapolation accuracy.
Denote the output of coarse extrapolation subnetwork as $\hat{\mathbf{u}}_{\mathcal{A}}^{c}$  and the output of fine extrapolation subnetwork as $\hat{\mathbf{u}}_{\mathcal{A}}^{f}$.
We propose to formulate the fine extrapolation (from $\hat{\mathbf{u}}_{\mathcal{A}}^{c}$ to $\hat{\mathbf{u}}_{\mathcal{A}}^{f}$) as an optimization problem that satisfies the following ODE
\begin{equation}\label{ode_initial}
\left\{
\begin{aligned}
&\frac{d\mathbf{u}_{f}(t)}{dt}=\bm\psi\left[\mathbf{u}_{f}(t),t\right]\\
&\mathbf{u}_{f}(t_{0})=\hat{\mathbf{u}}_{\mathcal{A}}^{c}\\
\end{aligned}
\right.,
\end{equation}
where $\mathbf{u}_{f}(\cdot)$ denotes the fine extrapolation function, and $\bm\psi(\cdot)$ denotes the differential function of $\mathbf{u}_{f}(t)$.
The initial condition $\mathbf{u}_{f}(t_{0})$ is $\hat{\mathbf{u}}_{\mathcal{A}}^{c}$, while the final condition $\mathbf{u}_{f}(t_{N})$ is  $\hat{\mathbf{u}}_{\mathcal{A}}^{f}$.
The derivation process can be found in  Appendix \ref{appendix_ode}.
Traditional methods to solve ODEs are Runge-Kutta \cite{dormand1980family} and Multi-step methods \cite{gragg1964generalized}. With the precise knowledge of $\bm\psi(\cdot)$, there is
\begin{equation}\label{ode_solver}
\mathbf{u}_{f}(t_{N})=\mathbf{u}_{f}(t_{0})+\int_{t_{0}}^{t_{N}}\frac{d\mathbf{u}_{f}(t)}{dt}dt=\mathbf{u}_{f}(t_{0})+\int_{t_{0}}^{t_{N}}\bm\psi\left[\mathbf{u}_{f}(t),t\right].
\end{equation}
However,  since $\bm\psi(\cdot)$ is not available in the considered extrapolation, we could not use  (\ref{ode_solver}) to solve (\ref{ode_initial}).
We then design  the ADEN by combining the deep neural network and the structure of Runge-Kutta solution (in \cite{dormand1980family}) whose structure is shown in  Fig. \ref{ADEN} and the mathematical expression is formulated as
\begin{equation}\label{RK}
\begin{aligned}
\mathscr{K}_{0}&=f_{0}\left(\hat{\mathbf{u}}_{\mathcal{A}}^{c}\right),\quad &\mathscr{K}_{1}&=f_{\mathscr{K}_{1}}\left(\mathscr{K}_{0}\right),\\
\mathscr{K}_{2}&=f_{\mathscr{K}_{2}}\left(\mathscr{K}_{0}+a_{1}\mathscr{K}_{1}\right),\quad &\mathscr{K}_{3}&=f_{\mathscr{K}_{3}}\left(\mathscr{K}_{0}+a_{2}\mathscr{K}_{2}\right),\\
\mathscr{K}_{4}&=f_{\mathscr{K}_{4}}\left(\mathscr{K}_{0}+a_{3}\mathscr{K}_{3}\right),\quad &\hat{\mathbf{u}}_{\mathcal{A}}^{f}&=f_{5}\left(\mathscr{K}_{0}+b_{1}\mathscr{K}_{1}+b_{2}\mathscr{K}_{2}+b_{3}\mathscr{K}_{3}+b_{4}\mathscr{K}_{4}\right),\\
\end{aligned}
\end{equation}
where $f_{\mathscr{K}_{i}}(\cdot)$ is the nonlinear mapping (or function) of sub-network $\mathscr{K}_{i}$, and $a_{i}, b_{i}$ are the parameters that will be trained.

The penalty of ADEN is set as
\begin{equation}\label{penalty_ode}
\mathcal{L}_{ADEN}=\beta_{1}\rVert \mathbf{u}_{\mathcal{A}}-\hat{\mathbf{u}}_{\mathcal{A}}^{c}\rVert_{2}^{2}+\beta_{2}\rVert \mathbf{u}_{\mathcal{A}}-\hat{\mathbf{u}}_{\mathcal{A}}^{f}\rVert_{2}^{2},
\end{equation}
where $\beta_{1}\textgreater0$ and $\beta_{2}\textgreater0$ are the tuning parameters of the two parts of penalties.

\subsection{Joint Training of ASN and ADEN}
\begin{algorithm}[t]
\setstretch{1.8}
\caption{Joint Training of ASN and ADEN}
\label{alg1}
\begin{algorithmic}
\REQUIRE Training dataset $\mathcal{D}$, Number of iterations $n_{iter}$, hyperparameters $\alpha_{1},\,\alpha_{2},\,\rho$, initialized trainable parameters $\bm{\Theta_{s}}$, $\bm{\Theta_{e}}$ and $a_{1}=a_{2}=\frac{1}{2},a_{3}=1,b_{1}=b_{4}=\frac{1}{6},b_{2}=b_{3}=\frac{1}{3}$.
\ENSURE Trained ASN parameters $\bm{\Theta_{s}}$, antenna selecting vector $\mathbf{s}$ and AEN parameters $\bm{\Theta_{e}}$.
\FOR{i = k to $n_{iter}$}
\STATE ASN Phase:
\STATE - Draw mini-batch $\mathcal{D}_{k}$: a random subset of $\mathcal{D}$
\STATE - Generate antenna selection vector $\mathbf{s}$ by: $\mathbf{s}=\rm{M_{t}\text{-}hot}(\mathop{\rm arg\,top\,M_{t}}\limits_{i}\left\{p_{i}|i=1,2,\cdots,N_{t}\right\})$
\STATE - Generate a differentiable approximation of $\mathbf{s}$: $\widetilde{\mathbf{p}}=M_{t}\cdot\mathbf{p}$
\STATE - Separate the real and imaginary parts of $\mathbf{u}_{\mathcal{A}}$:\;$\mathbf{Z}_{in}\triangleq[\Re(\mathbf{u}_{\mathcal{A}}),\Im(\mathbf{u}_{\mathcal{A}})]$
\STATE - Concatenate $\mathbf{s}$ for real and imaginary parts of $\mathbf{u}_{\mathcal{A}}$ (same operation for $\widetilde{\mathbf{p}}$): $\overline{\mathbf{s}}=[\mathbf{s}^{T},\mathbf{s}^{T}]^{T}$
\STATE - Perform antenna selecting (Hadamard product) (same operation for $\widetilde{\mathbf{p}})$: $\mathbf{Z}_{s}\triangleq\overline{\mathbf{s}}\odot\mathbf{Z}_{in}$
\STATE AEN Phase:
\STATE - Compute the output of extrapolation network: $\mathbf{Z}_{out}\triangleq f\left(\mathbf{Z}_{s},\bm{\Theta}_{s}\right)$
\STATE - Compute the loss function: $\mathcal{L}$
\STATE Back-propagation Phase:
\STATE - Replace $\mathbf{s}$ with $\widetilde{\mathbf{p}}$
\STATE - Use Adam optimizer to update $\bm{\Theta_{s}}$ and $\bm{\Theta_{e}}$
\ENDFOR
\end{algorithmic}
\end{algorithm}

We adopt a combined loss function to jointly train ASN and ADEN:
\begin{equation}\label{loss-function}
\mathcal{L}=\mathcal{L}_{ASN}+\rho\mathcal{L}_{ADEN},
\end{equation}
where $\bm\rho$ is the weight to balance the penalties of ASN and ADEN.

The detailed steps of the joint training algorithm are described in Algorithm \ref{alg1}.
Since the neural network can only process real numbers, we construct the input $\mathbf{u}_{\mathcal{A}}$ (in Fig. \ref{jointasnaen}) as
\begin{equation}\label{Z_in}
\mathbf{Z}_{in}\triangleq[\Re(\mathbf{u}_{\mathcal{A}}),\Im(\mathbf{u}_{\mathcal{A}})].
\end{equation}
Correspondingly, the antenna selection vector should also be constructed as
\begin{equation}\label{s_in}
\overline{\mathbf{s}}\triangleq[\mathbf{s}^{T},\mathbf{s}^{T}]^{T}.
\end{equation}
Then the input of ADEN is $\mathbf{Z}_{s}\triangleq\overline{\mathbf{s}}\odot\mathbf{Z}_{in}$.
The output of ADEN represents the real part and imaginary part of extrapolated $\hat{\mathbf{u}}_{\mathcal{A}}$ (in Fig. \ref{jointasnaen})
\begin{equation}\label{Z_out}
\mathbf{Z}_{out}\triangleq[\Re(\hat{\mathbf{u}}_{\mathcal{A}}),\Im(\hat{\mathbf{u}}_{\mathcal{A}})].
\end{equation}

After joint training of ASN and ADEN, we  obtain an antenna selection vector $\mathbf{s}$.
During the online evaluation, since the antenna selection vector $\mathbf{s}$ has been obtained, we can delete the ASN and use $\mathbf{s}\odot\mathbf{u}_{\mathcal{A}}$ for antenna domain extrapolation.

\section{Typical Applications in Transceiver Design}\label{section4}
The proposed antenna selection can be applied into many communications tasks that need  to select antennas with certain purpose, for example, antenna selection for the channel estimation in hybrid massive MIMO system, antenna activation strategy for RIS, antenna selection for data transmission, etc.
Similarly, the antenna domain extrapolation can be applied in many extrapolation problems in multi-antenna systems.
Next, we explain how our proposed model can be applied to the channel, beam, and covariance extrapolation problems.

\subsection{Channel Extrapolation}
In the channel extrapolation case, we have $\mathbf{u}_{\mathcal{A}}=\mathbf{h}_{\mathcal{A}}$  and $\mathbf{u}_{\mathcal{B}}=\mathbf{h}_{\mathcal{B}}$   in (\ref{mapping}).
The channel extrapolation can be described as
\begin{equation}\label{h_extrapolation}
\begin{aligned}
\mathbf{h}_{\mathcal{A}} = f_{h}\left(\mathbf{h}_{\mathcal{B}}\right),
\end{aligned}
\end{equation}
where $f_{h}(\cdot)$ is the channel extrapolation function learned by the ADEN.

We set the network input as $\mathbf{h}_{in}=\left[\Re\left(\mathbf{h}^{T}_{\mathcal{B}}\right),\Im\left(\mathbf{h}^{T}_{\mathcal{B}}\right)\right]^{T}$,
and the corresponding label is $\mathbf{h}_{lab}=\left[\Re\left(\mathbf{h}^{T}_{\mathcal{A}}\right),\Im\left(\mathbf{h}^{T}_{\mathcal{A}}\right)\right]^{T}.$

The quality of the channel extrapolation result is evaluated by the NMSE indicator.
\begin{equation}\label{nmse_define}
\mathbf{NMSE} = \frac{\mathbb{E}\left[\left|\mathbf{h}_{\mathcal{A}}-\hat{\mathbf{h}}_{\mathcal{A}}\right|^{2}\right]}{\mathbb{E}\left[\left|\mathbf{h}_{\mathcal{A}}\right|^{2}\right]}.
\end{equation}

\subsection{Beam Prediction}
The beamforming  is used  to increase  downlink transmission rate of massive MIMO system, and the optimal  downlink beamforming vector $\mathbf{f}_{\mathcal{A}}$ can be generated from (\ref{beam}). When there can only be  limited number of pilots and only limited channel $\mathbf{h}_{\mathcal{B}}$ of a small number of antennas can be obtained, we propose to directly predict the beam index $\mathbf{Beam}_{\mathcal{A}}$ from $\mathbf{h}_{\mathcal{B}}$. In this case,  we set $\mathbf{u}_{\mathcal{A}}=\mathbf{Beam}_{\mathcal{A}}$  and $\mathbf{u}_{\mathcal{B}}=\mathbf{h}_{\mathcal{B}}$   in (\ref{mapping}). The mathematical formula of beam prediction is
\begin{equation}\label{b_extrapolation}
\left\{\mathbf{Beam}_{\mathcal{A}}\right\} = f_{b}(\left\{\mathbf{h}_{\mathcal{B}}\right\}),
\end{equation}
where $f_{b}(\cdot)$ is the beam extrapolation function that will be learned by the ADEN.
After the $\mathbf{Beam}_{\mathcal{A}}$ being predicted, the corresponding beamforming vector can be obtained by looking up the codebook.
\subsection{ CCM Extrapolation}

CCM is the statistical characteristic of the channel and is conventionally obtained from the accumulation of sufficient number of estimated channel vectors.
For massive MIMO system, unfortunately, the number of the estimated channel vectors is significantly  large.
Nevertheless, according to the mapping (\ref{mapping-ccm}),  CCM can also be extrapolated from a small number of antennas.

Let $\mathbf{u}_{\mathcal{A}}=\mathbf{R}_{\mathcal{A}}$   and $\mathbf{u}_{\mathcal{B}}=\mathbf{R}_{\mathcal{B}}$ in (\ref{mapping}), and there is
\begin{equation}\label{c_extrapolation}
\mathbf{R}_{\mathcal{A}} = f_{c}\left(\mathbf{R}_{\mathcal{B}}\right),
\end{equation}
where $f_{c}(\cdot)$ denotes the CCM extrapolation function that will be learned by the ADEN.
Similarly to the channel vector extrapolation, the input of CCM extrapolation network is $\mathbf{R}_{in}=\left\{\Re\left(\mathbf{R}_{\mathcal{B}}\right),\Im\left(\mathbf{R}_{\mathcal{B}}\right)\right\}$,
and the corresponding label is
$\mathbf{R}_{lab}=\left\{\Re\left(\mathbf{R}_{\mathcal{A}}\right),\Im\left(\mathbf{R}_{\mathcal{A}}\right)\right\}$.
Denote $\mathbf{R}_{out}=\{\Re(\mathbf{R}_{0}),\Im(\mathbf{R}_{0})\}$ as the output of ADEN.
Since the CCMs are positive semi-definite, we add one more layer to ensure the positive semi-definiteness, and the corresponding output is
\begin{equation}\label{semi-definite}
\mathbf{R}_{out}^{'}=\{\Re(\mathbf{R}_{0})+\Re(\mathbf{R}_{0}^{T}),\Im(\mathbf{R}_{0})-\Im(\mathbf{R}_{0}^{T})\}=\{\Re(\mathbf{R}_{0}+\mathbf{R}_{0}^{H}),\Im(\mathbf{R}_{0}+\mathbf{R}_{0}^{H})\}.
\end{equation}

\section{Simulation Result}\label{section5}
In this section, we evaluate the performance of the proposed ASN and ADEN for channel extrapolation, beam prediction, and CCM extrapolation.

\subsection{Communications Set Up}
\begin{figure*}[t]
\centering
\includegraphics[width=0.45\textwidth]{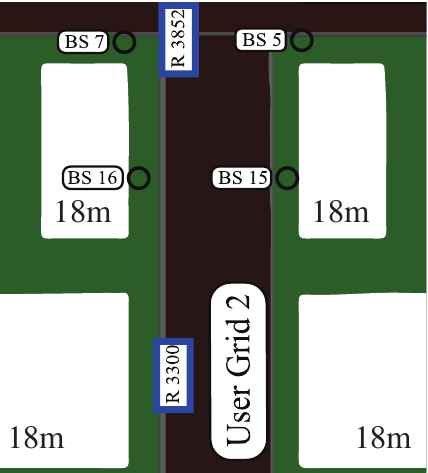}
\caption{The communications scenario.}
\label{deepmimo}
\end{figure*}
Let us consider a scenario from the DeepMIMO dataset \cite{deepMIMOdataset} that is constructed from the 3D ray-tracing software Wireless InSite \cite{WinNT} and could  capture the channel dependence on the frequency and location.
Specifically, we use the outdoor scenario `O1\_28' \cite{deepMIMOdataset} available at frequency $f_{c}=28$ GHz, as shown in Fig. \ref{deepmimo}.
Meanwhile, the BS (BS 15 in Fig. \ref{deepmimo}) is equipped with a uniform planar array (UPA) of 8$\times$8 antennas while the user has only one antenna.
The antenna spacing $d$ is set to $\frac{\lambda_{c}}{2}$ where $\lambda_{c}$ is the carrier wavelength.
The bandwidth of the system is set to 200 MHz and the number of paths is set to 11.
The corresponding rows of the communication scenario in Fig. \ref{deepmimo}  are from 3252 to 3852.
Each row contains 181 users while each user represents a position in the scenario. Hence, there are a total number of 108,781 channels.
The channel vectors are generated based on formula (\ref{channelmodel}) and the parameters in Table~\ref{table1}.

\begin{table}[t]
\renewcommand\arraystretch{1.2}
\caption{DeepMIMO Dataset Parameters}\label{table1}
\centering
\begin{tabular}{p{8cm}<{\centering}|p{6cm}<{\centering}}
\hline
\hline
Parameters & Value  \\
\hline
Scenario name & O1\_28\\
\hline
Active BS & BS15\\
\hline
Active users & Row 3252-3852\\
\hline
Number of BS Antennas & 64\\
\hline
Number of BS Antennas in x-axis& 8\\
\hline
Number of BS Antennas in y-axis& 8\\
\hline
Number of BS Antennas in z-axis& 1\\
\hline
Antenna spacing (wave-length) & 0.5\\
\hline
Bandwidth (GHz) & 0.2\\
\hline
Number of OFDM subcarriers & 1\\
\hline
OFDM sampling factor & 1\\
\hline
OFDM limit & 1\\
\hline
Number of paths& 11\\
\hline
\hline
\end{tabular}
\end{table}

We use the algorithm \cite{beam_codebook} in \cite{2019arXiv191002900A} to generate a beamforming codebook based on the antenna array parameters in Table \ref{table1}.
Then we select the sequence number of the beamforming vector   according to formula (\ref{beam}) to form the label of the beam prediction network.

The dataset of CCM extrapolation is generated from the channel vectors that are collected from $1\times 1\,m^{2}$ area around each user.
Define one $1\times 1\,m^{2}$ area as a collection block.
For each collection block, we evenly collected channels at 25 locations i.e., 5 rows and 5 columns, and then obtain the channel covariance matrix as $\mathbf{R}_{\rm h} =\mathbb{E}\left[\mathbf{h}\cdot \mathbf{h}^{H}\right]$.
We collect the CCMs of the users from 3552 row to 2852 row and generate a total number of 52569 collection blocks.

\subsection{Neural Network Training}\label{training}
The configurations of the neural network in the three applications are as follows:
\subsubsection{Channel Extrapolation}
\begin{table*}[t]
\renewcommand\arraystretch{1.2}
\caption{Network Training Hyper-Parameters}\label{table2}
\centering
\begin{tabular}{m{5cm}<{\centering}|m{3cm}<{\centering}|m{3cm}<{\centering}|m{3cm}<{\centering}}
\hline
\hline
Parameters & Channel & Beam & CCM \\
\hline
Solver & \multicolumn{3}{c}{Adam}\\
\hline
Initial learning rate & \multicolumn{3}{c}{$1\times 10^{-3}$}\\
\hline
Sub-network $\mathscr{K}_{i}$ & \multicolumn{3}{c}{fully connected layer and ReLu}\\
\hline
Number of neurons in $\mathscr{K}_{i}$ & 512 & 512 &  16,386\\
\hline
Dataset size & 108,781 & 108,781 & 52,569\\
\hline
Dataset split & \multicolumn{3}{c}{80\%-20\%}\\
\hline
Penalty factor of vector $\mathbf{s}$ & \multicolumn{3}{c}{$\alpha_{1}=\alpha_{2}=\beta_{1}=1$, $\beta_{2}=10$}\\
\hline
Scale factor $\rho$ & \multicolumn{3}{c}{initial $\rho=5$, then $\times 5$ every epoch}\\
\hline
\hline
\end{tabular}
\end{table*}

In case of channel extrapolation, the ASN is composed of three layers.
Each layer has 64 neurons, and the output of ASN is $\mathbf{s}\in \left\{0,1\right\}^{64\times 1}$.
The input of ADEN is $\overline{\mathbf{s}}\odot \left[\Re(\mathbf{h}^{T}_{\mathcal{A}}),\Im(\mathbf{h}^{T}_{\mathcal{A}})\right]^{T}\in R^{128\times 1}$.
Moreover, each sub-network contains 512 neurons and a ReLu layer.
\subsubsection{Beam Prediction}
The input  is the same as that of the channel extrapolation case.
The output and label of the network is the beam index and the loss function is a crossentry function.
Moreover, each layer includes 128 neurons and the activation function is ReLu.
For beam prediction case, since the output of ADEN is the beam index, i.e., a one-dimension number, only a few layers are needed to  achieve satisfactory accuracy. Hence we delete the coarse extrapolation subnetwork and only use the fine extrapolation subnetwork for beam prediction.
\subsubsection{CCM Extrapolation}
In case of CCM extrapolation, the ASN is composed of three layers.
Each layer has 64 neurons, and the output of ASN is $\mathbf{s}\in \left\{0,1\right\}^{64\times 1}$.
Different from channel extrapolation, sampling the covariance matrix in the antenna domain requires expanding the antenna selection vector into a two-dimensional matrix $\mathbf{S}=\mathbf{s}\cdot\mathbf{s}^{T}\,\in\left\{0,1\right\}^{64\times64}$.
Then we concatenate $\mathbf{S}$ to generate $\overline{\mathbf{S}}=[\mathbf{S}^{T},\mathbf{S}^{T}]^{T}\in \left\{0,1\right\}^{128\times 64}$.
CCM should also be constructed as $\mathbf{R}_{in}\in R^{128\times64}$ with the real and imaginary parts separated.
Then we perform Hadamard product on $\overline{\mathbf{S}}$ and $\mathbf{Z}_{in}$ as $\overline{\mathbf{S}}\odot\left[\Re(\mathbf{R}^{T}_{\mathcal{A}}),\Im(\mathbf{R}^{T}_{\mathcal{A}})\right]^{T}$.
Before entering the ADEN network, the result of Hadamard product should be reshaped into a column vector $\in R^{8192\times 1}$.

\subsection{Performance Evaluation}

For all simulations, we compare four channel extrapolation schemes with the same number of neurons: (i) `Uniform + DNN' (using traditional DNN to extrapolate from uniform antenna selection patterns);
(ii) `Uniform + ADEN' (using ADEN to extrapolate from uniform antenna selection patterns);
(iii) `ASN + DNN' (using DNN to extrapolate from learned antenna selection patterns);
(iv) `ASN + ADEN' (using ADEN to extrapolate from learned antenna selection patterns).

\subsubsection{Channel Extrapolation}

\begin{figure}[t]
\centering
\subfigure[]{
\begin{minipage}[t]{0.32\linewidth}
\centering
\includegraphics[width=2.02in]{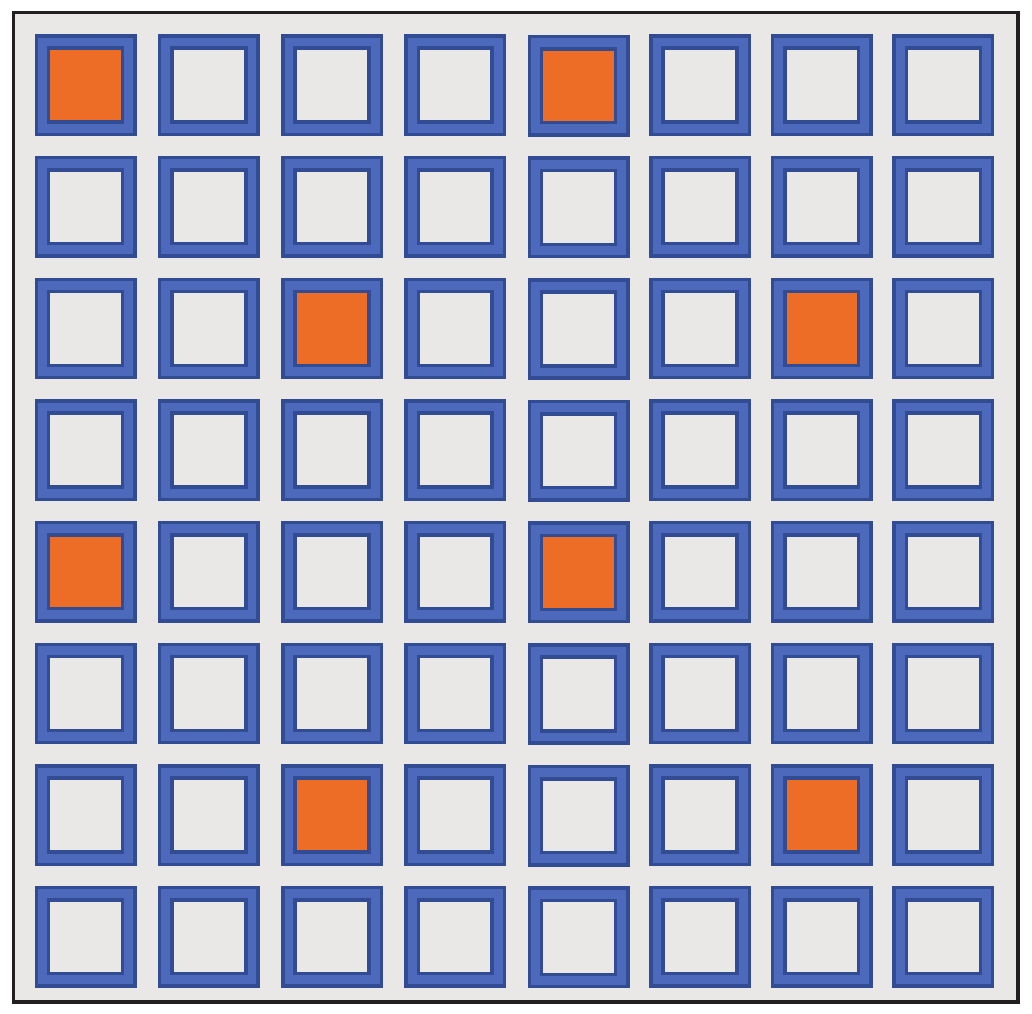}
\end{minipage}%
}%
\subfigure[]{
\begin{minipage}[t]{0.32\linewidth}
\centering
\includegraphics[width=2in]{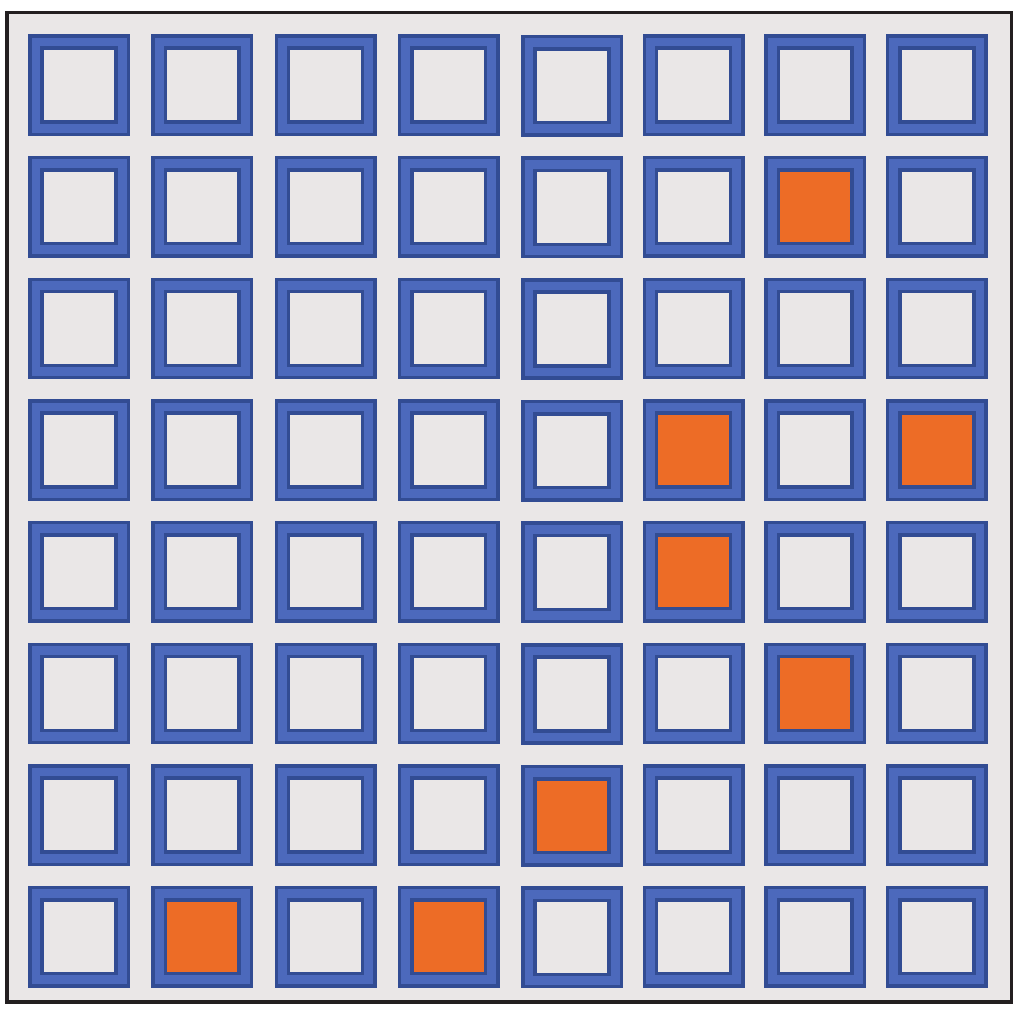}
\end{minipage}%
}%
\subfigure[]{
\begin{minipage}[t]{0.32\linewidth}
\centering
\includegraphics[width=2in]{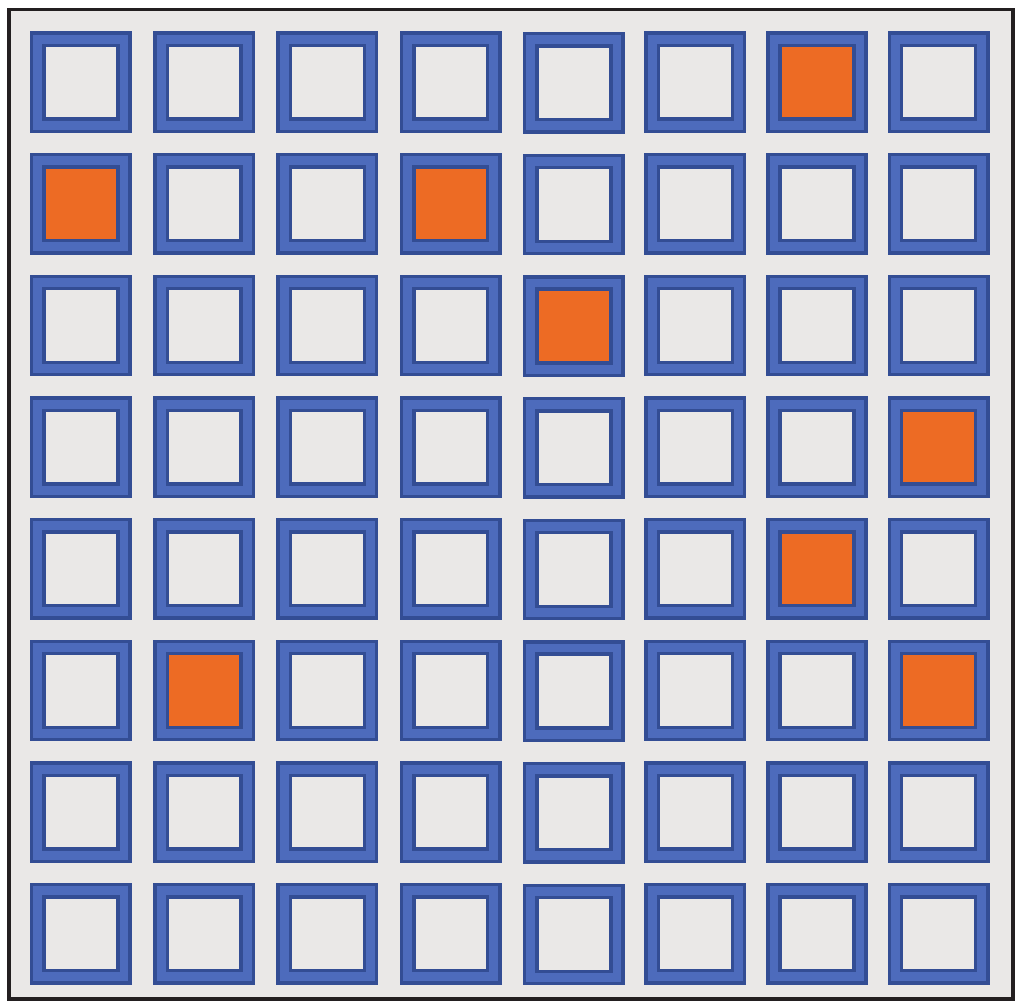}
\end{minipage}
}%
\centering
\caption{Antenna selection patterns at SNR=30dB: (a) uniform antenna selection pattern; (b) antenna selection pattern learned by `ASN+DNN'; (c) antenna selection patterns learned by `ASN+ADEN'. }
\label{channel-pattern}
\end{figure}
\begin{figure}[t]
\centering
\includegraphics[width=0.6\textwidth]{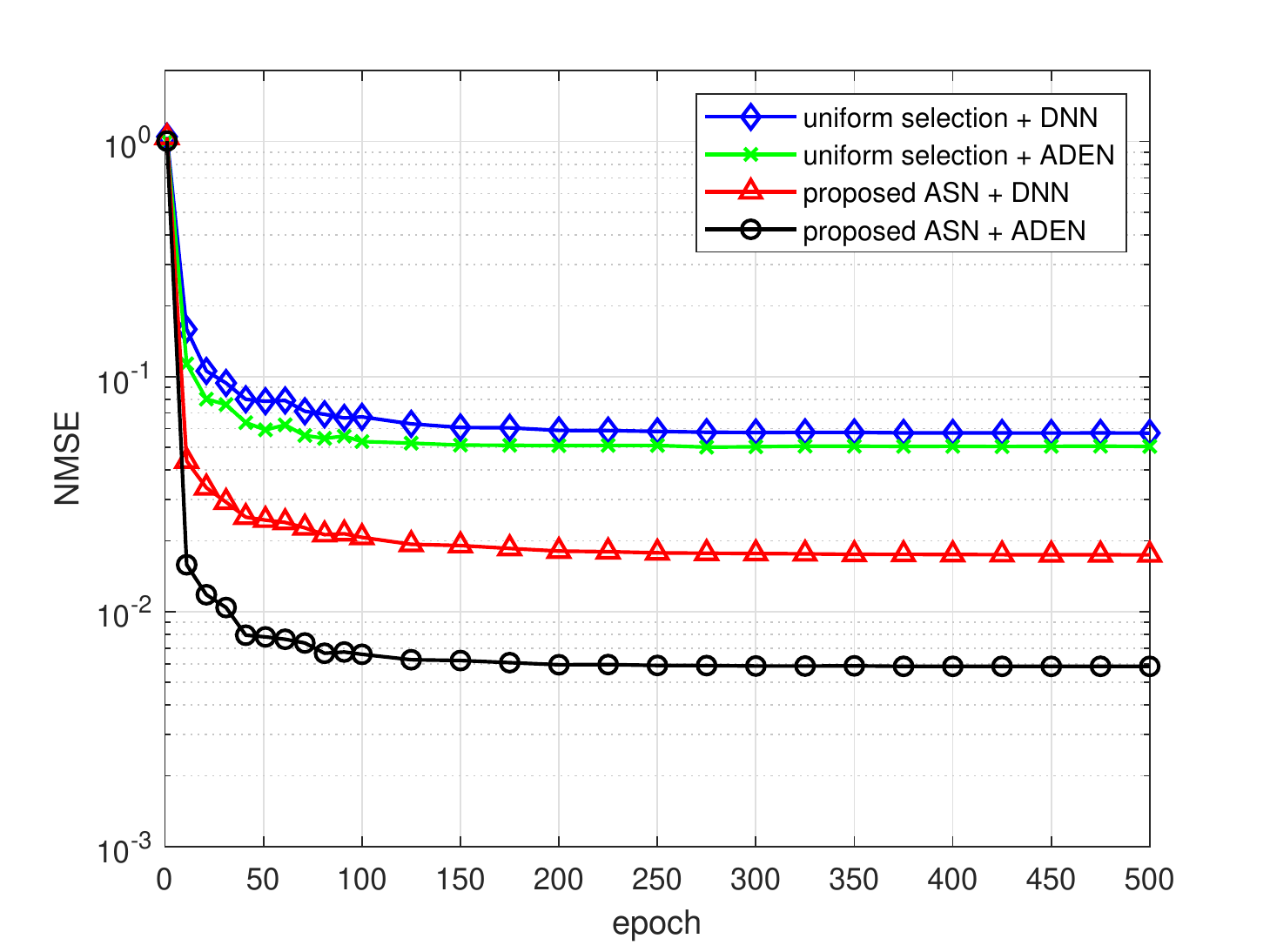}
\caption{The NMSE of channel extrapolation versus epoches with 8 antennas and SNR=30dB.}
\label{channel-epoch}
\end{figure}

\begin{figure}[t]
\centering
\includegraphics[width=0.6\textwidth]{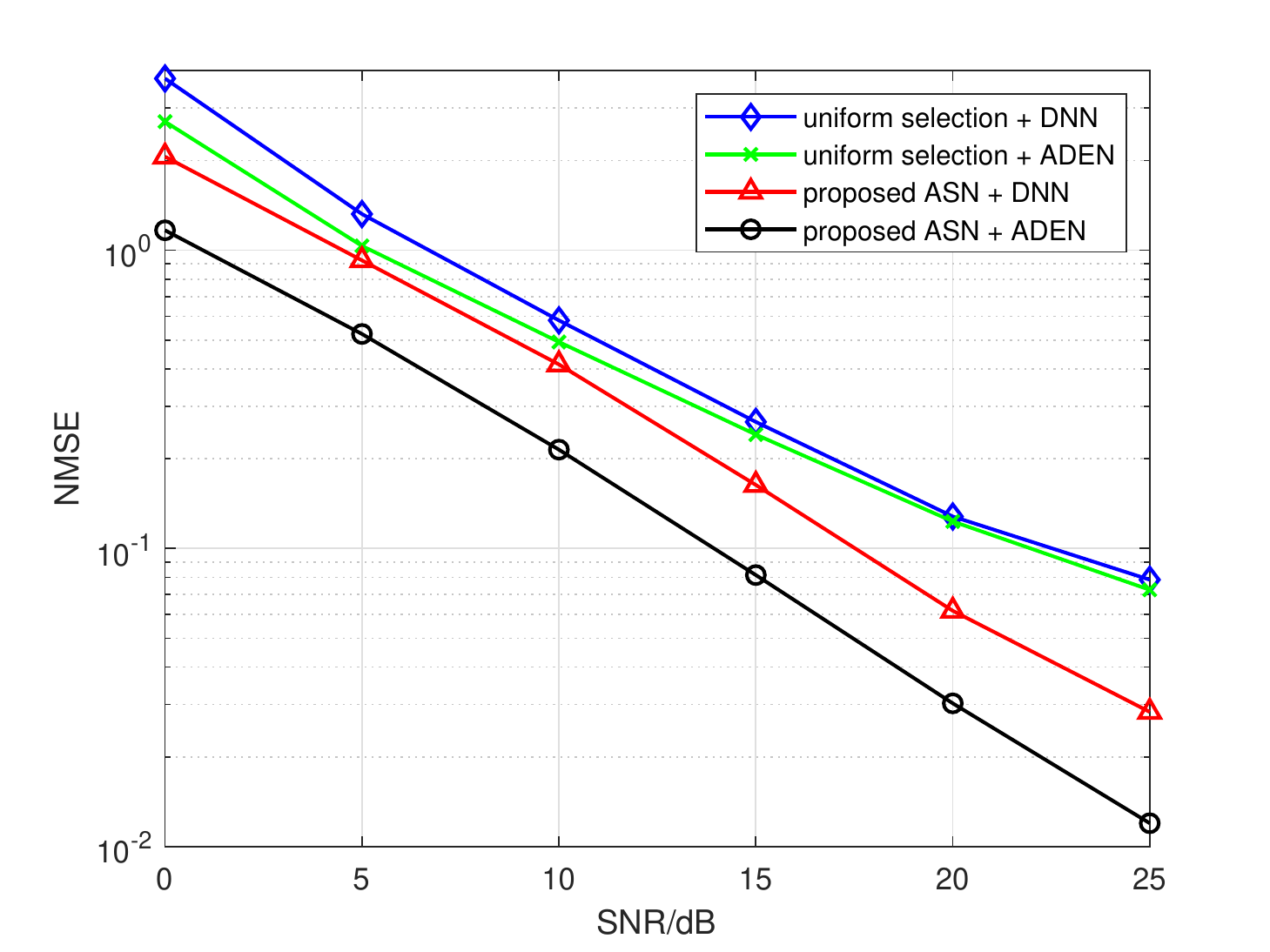}
\caption{The NMSE of channel extrapolation versus SNR with 8 antennas.}
\label{channel-snr}
\end{figure}

We use the channel of 8 antennas to extrapolate the channel of 64 antennas.
The uniform antenna selection pattern is shown in Fig. \ref{channel-pattern}(a).
At SNR=30dB, the antenna selection patterns learned by `ASN + DNN' and `ASN + ADEN' are shown in Fig.~\ref{channel-pattern}(b), Fig.~(c) respectively, which
look quite different from the uniform one. The NMSE of channel extrapolation versus the number of epochs  for four different schemes are displayed in Fig. \ref{channel-epoch}.
It is seen that the extrapolation NMSE of `Uniform + DNN'  is 0.060, while the extrapolation NMSE of `Uniform + ADEN' reaches 0.050.
Moreover, the extrapolation NMSE of `ASN + DNN' reduces to 0.017, while the extrapolation NMSE of `ASN + ADEN' significantly drops to 0.006.
Then we test the channel extrapolation NMSE at different SNR in Fig. \ref{channel-snr}.
Clearly, ADEN   outperforms traditional DNN in terms of the accuracy of extrapolation, and the proposed ASN performs much better than the uniform selection.
From Fig .\ref{channel-epoch} and Fig. \ref{channel-snr}, the proposed ADEN performs slightly better than DNN for uniform extrapolation. However, with the optimized antenna selection, the accuracy of ADEN will be much better than that of DNN, which demonstrates the effectiveness of the proposed joint training  scheme.

\subsubsection{Beam Prediction}
\begin{figure}[t]
\centering
\subfigure[]{
\begin{minipage}[t]{0.45\linewidth}
\centering
\includegraphics[width=2.5in]{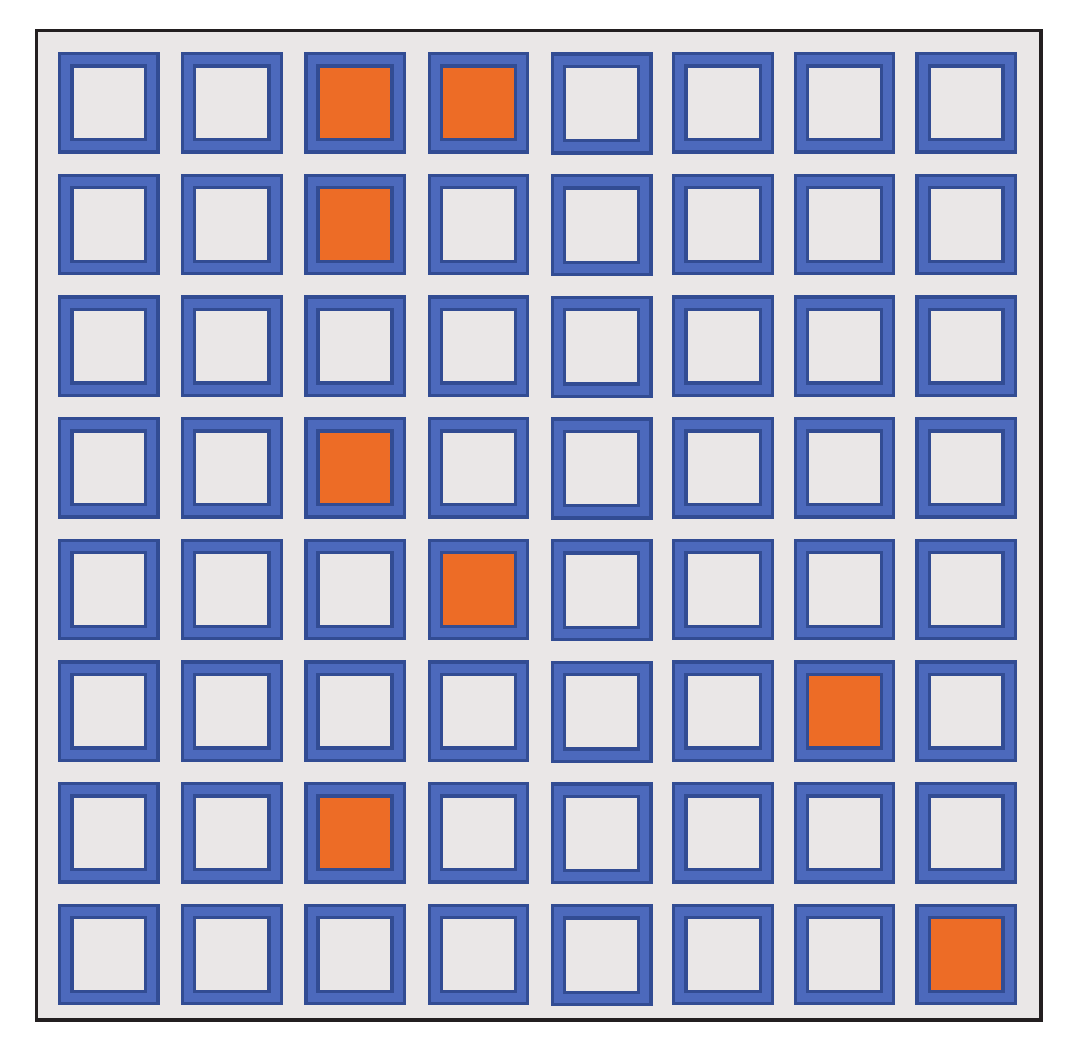}
\end{minipage}%
}%
\subfigure[]{
\begin{minipage}[t]{0.45\linewidth}
\centering
\includegraphics[width=2.5in]{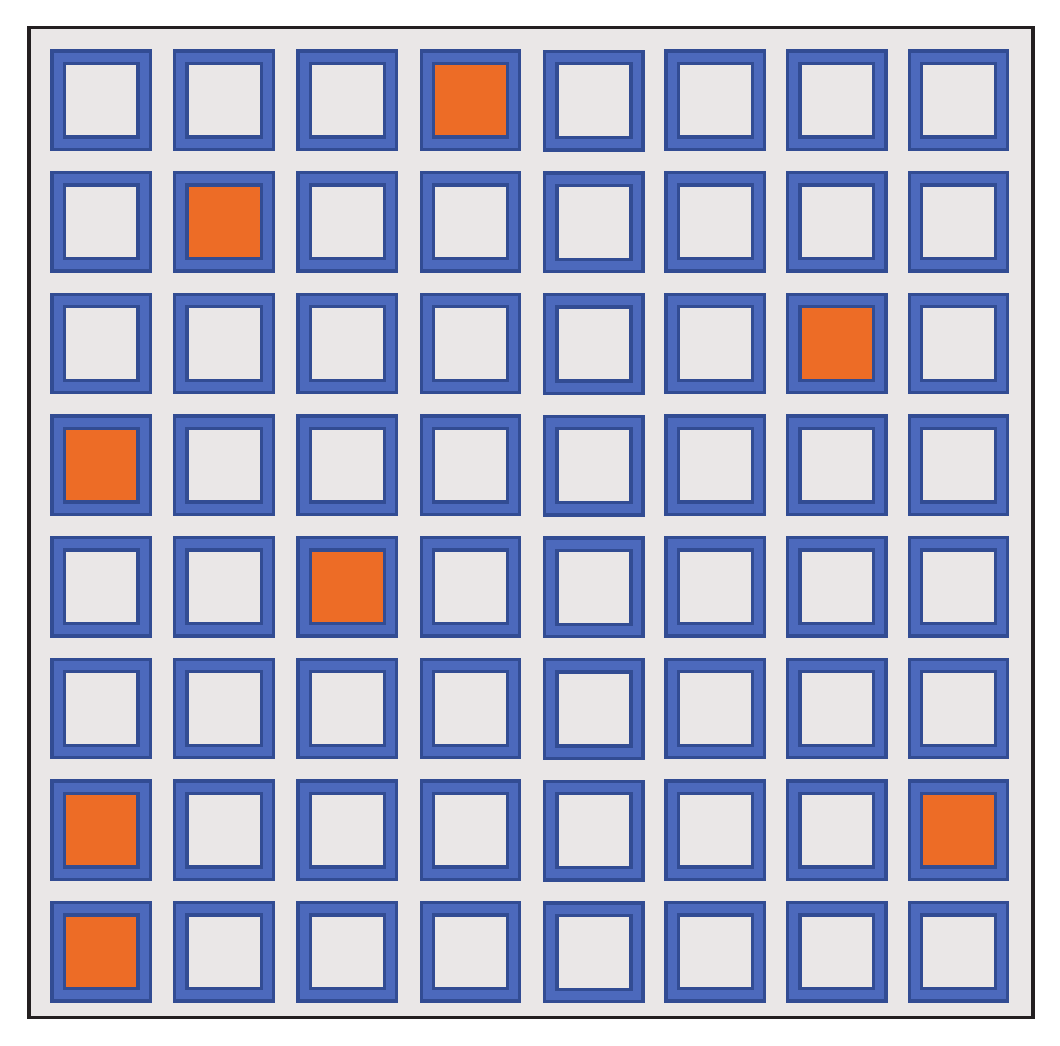}
\end{minipage}
}%
\centering
\caption{Antenna selection patterns at SNR=30dB: (a) antenna selection pattern learned by `ASN+DNN'; (b) antenna selection patterns learned by `ASN+ADEN'.}
\label{beam-pattern}
\end{figure}

\begin{figure}[t]
\centering
\includegraphics[width=0.6\textwidth]{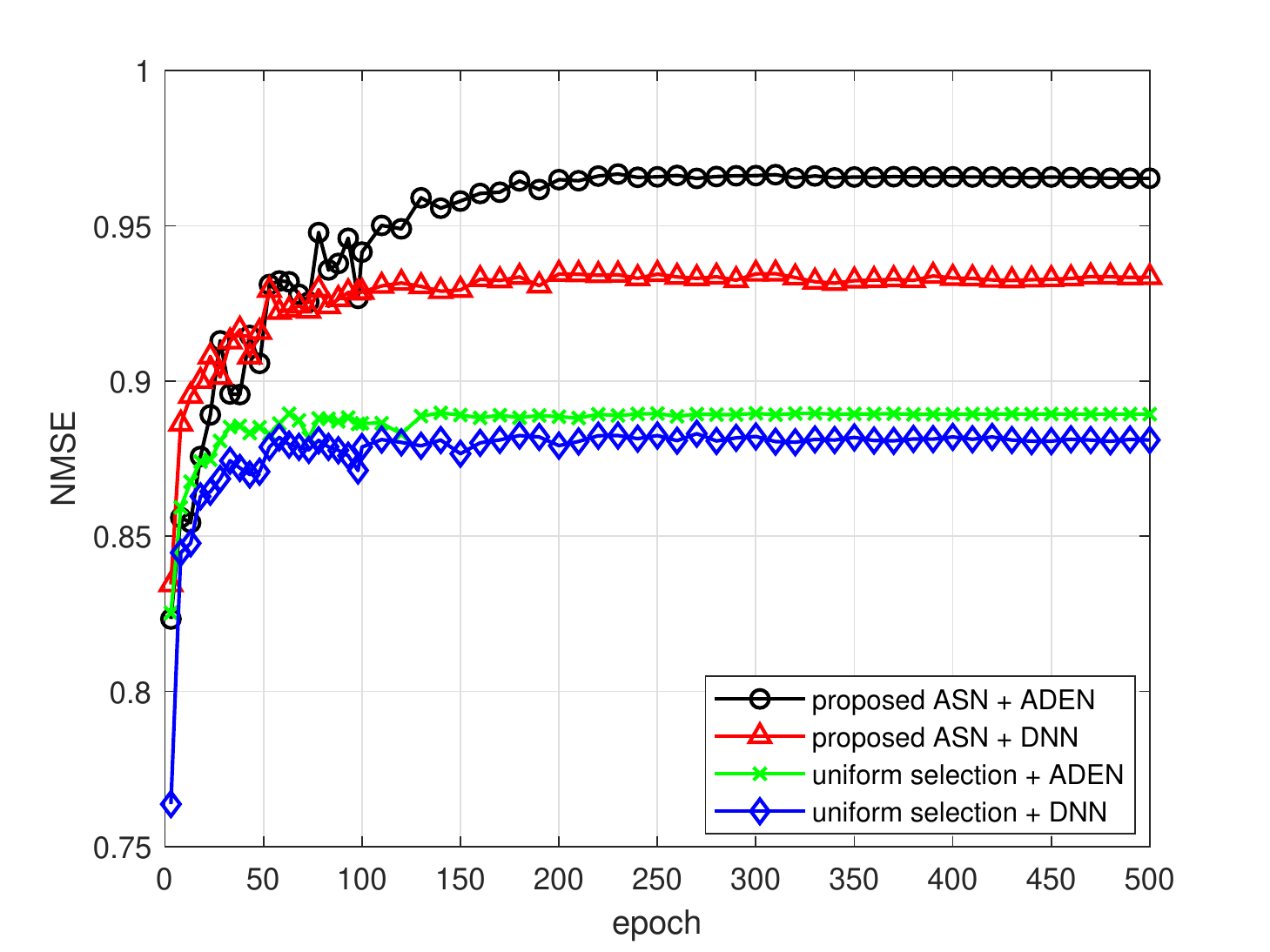}
\caption{Beam prediction accuracy versus epoches at SNR=30dB.}
\label{beam-epoch}
\end{figure}
\begin{figure}[t]
\centering
\includegraphics[width=0.6\textwidth]{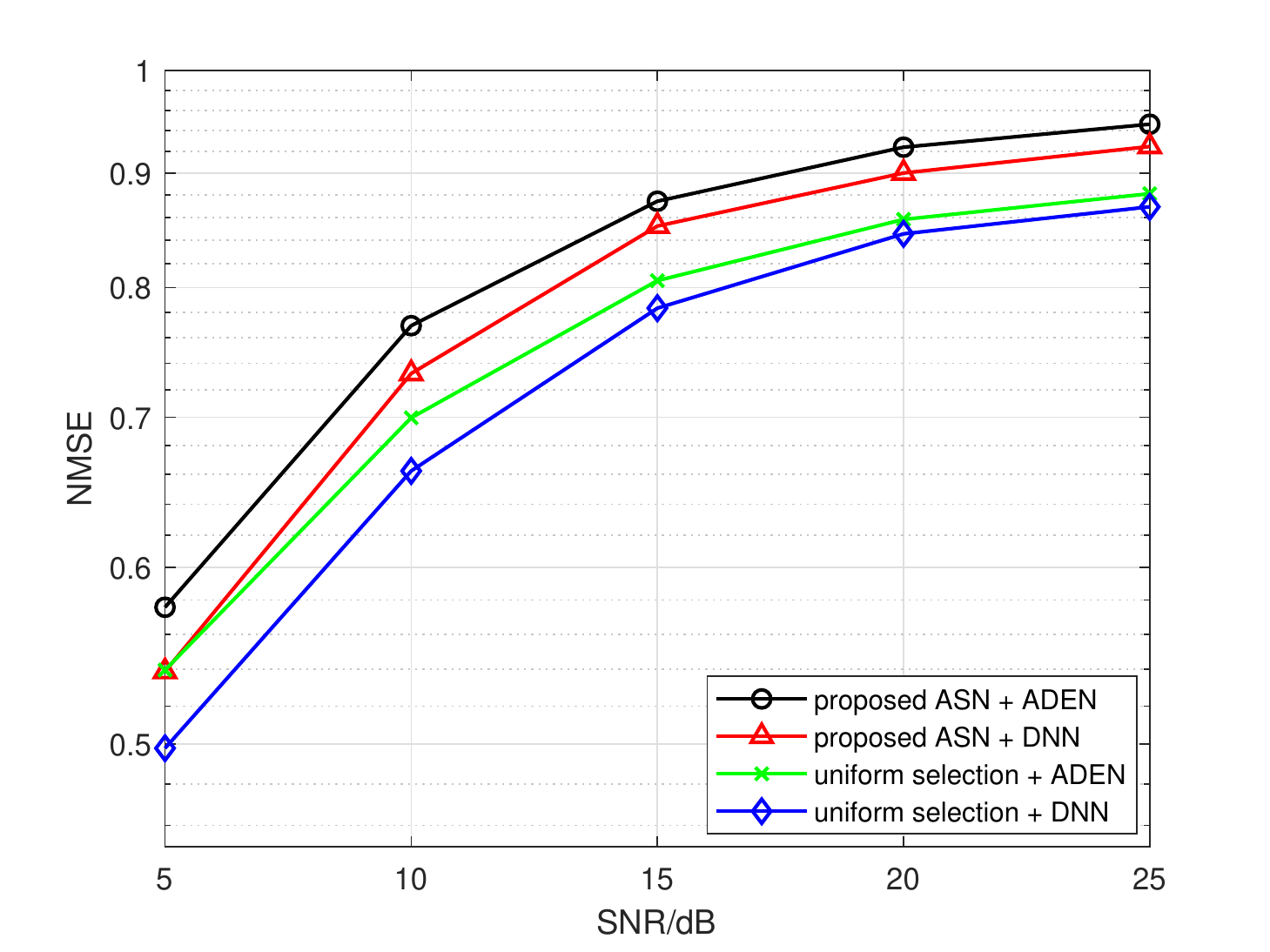}
\caption{Beam prediction accuracy versus SNR.}
\label{beam-SNR}
\end{figure}

For beam prediction, we utilize the channel of 8 antennas to predict the beam index of 64 antennas. At SNR=30dB, the antenna selection patterns learned by `ASN + DNN' and `ASN + ADEN' are shown in Fig. \ref{beam-pattern}.
The beam prediction accuracy of the four schemes are displayed in Fig. \ref{beam-epoch}.
The accuracy of `ASN + ADEN', `ASN + DNN', `Uniform + ADEN', and `Uniform + DNN' are 0.965, 0.933, 0.890, and 0.880 respectively.
Then we test the accuracy of beam prediction at different SNR in Fig. \ref{beam-SNR}.
Similarly to the channel extrapolation, we see  that the ADEN achieves high beam prediction accuracy than traditional DNN and the learned antenna selection pattern by ASN is better than the uniform  pattern.
\subsubsection{CCM Extrapolation}
\begin{figure}[t]
\centering
\subfigure[]{
\begin{minipage}{7cm}
\centering
\includegraphics[scale=0.55]{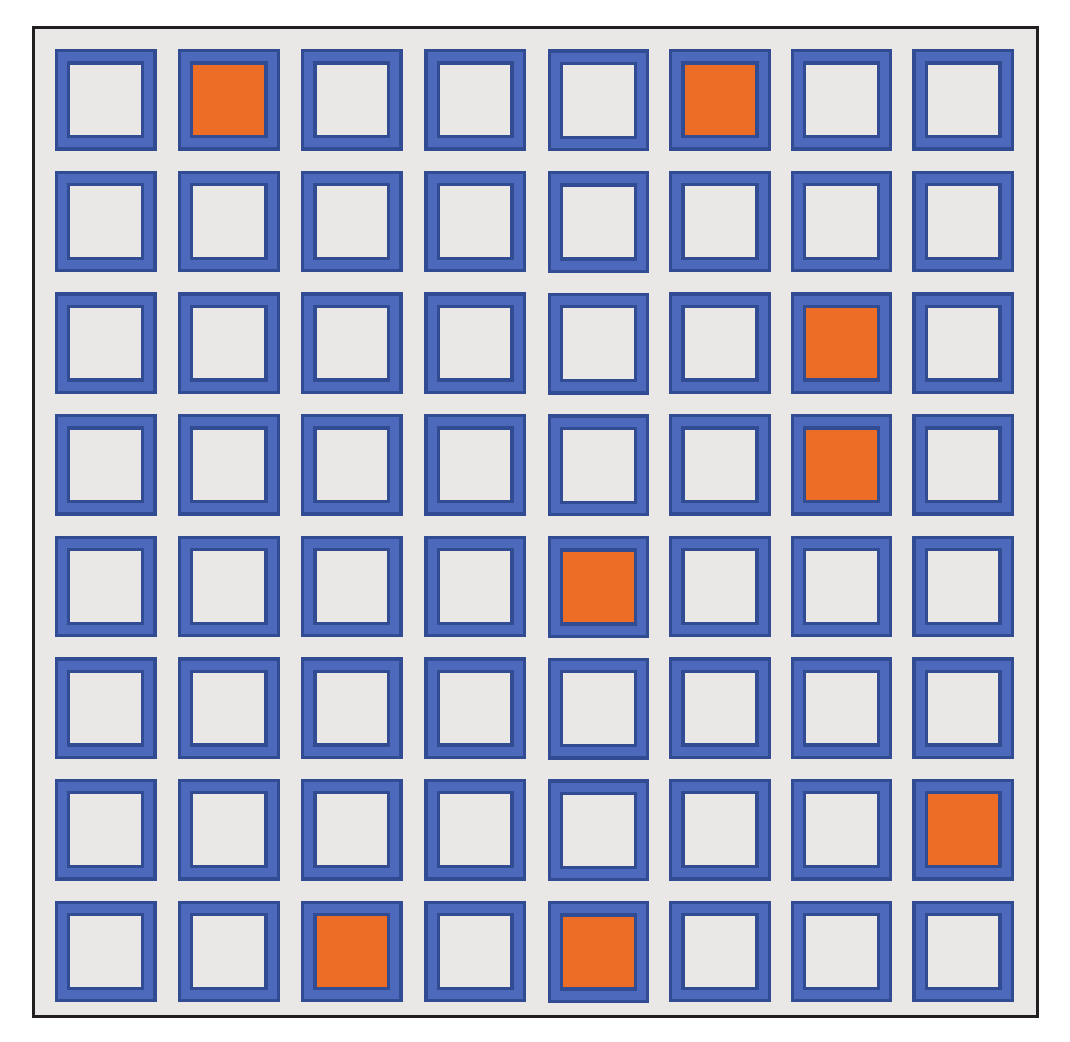}
\end{minipage}
}
\subfigure[]{
\begin{minipage}{7cm}
\centering
\includegraphics[scale=0.55]{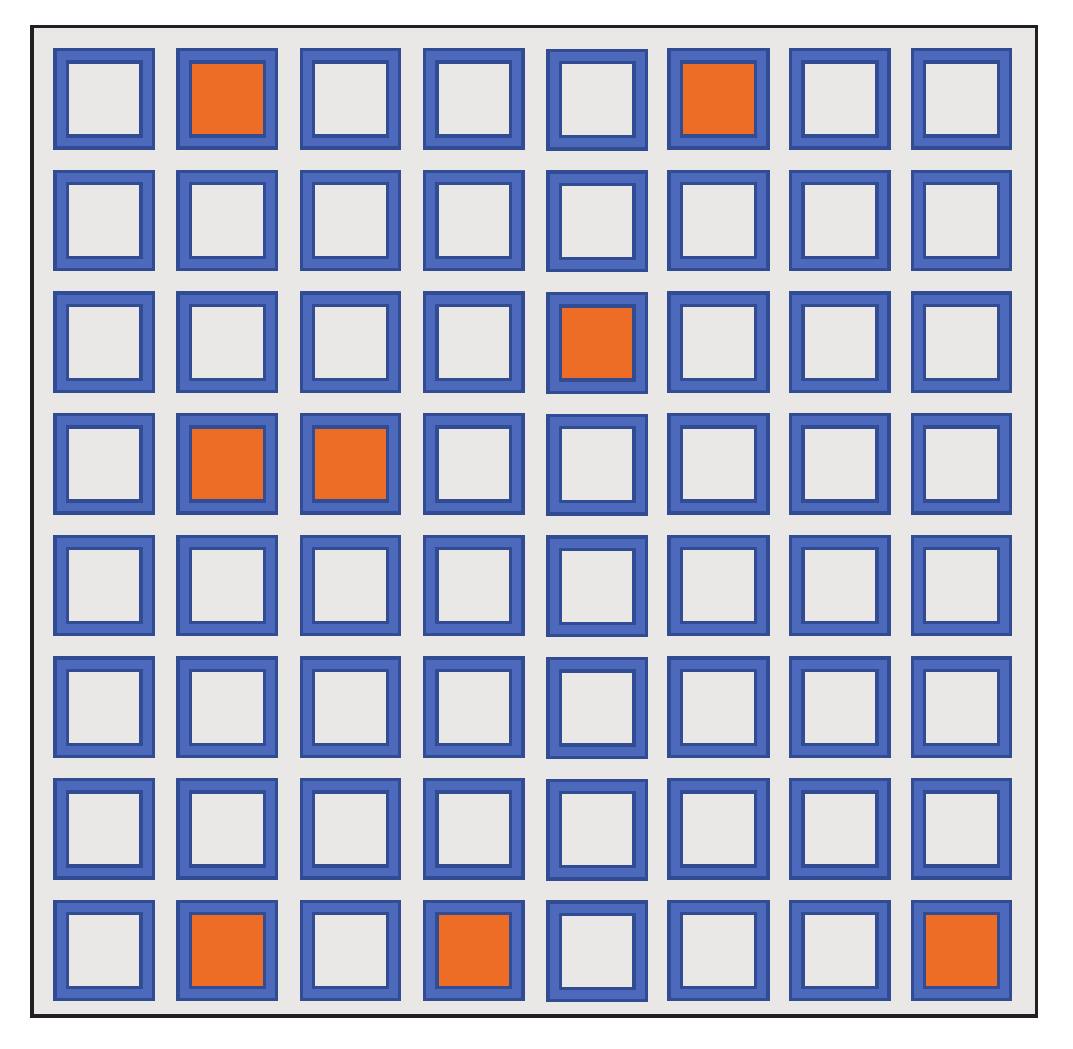}
\end{minipage}
}
\caption{Antenna selection patterns at SNR=30dB: (a) antenna selection pattern learned by `ASN+DNN'; (b) antenna selection pattern learned by `ASN+ADEN'.}
\label{ccm-pattern}
\end{figure}

\begin{figure}[t]
\centering
\includegraphics[width=0.6\textwidth]{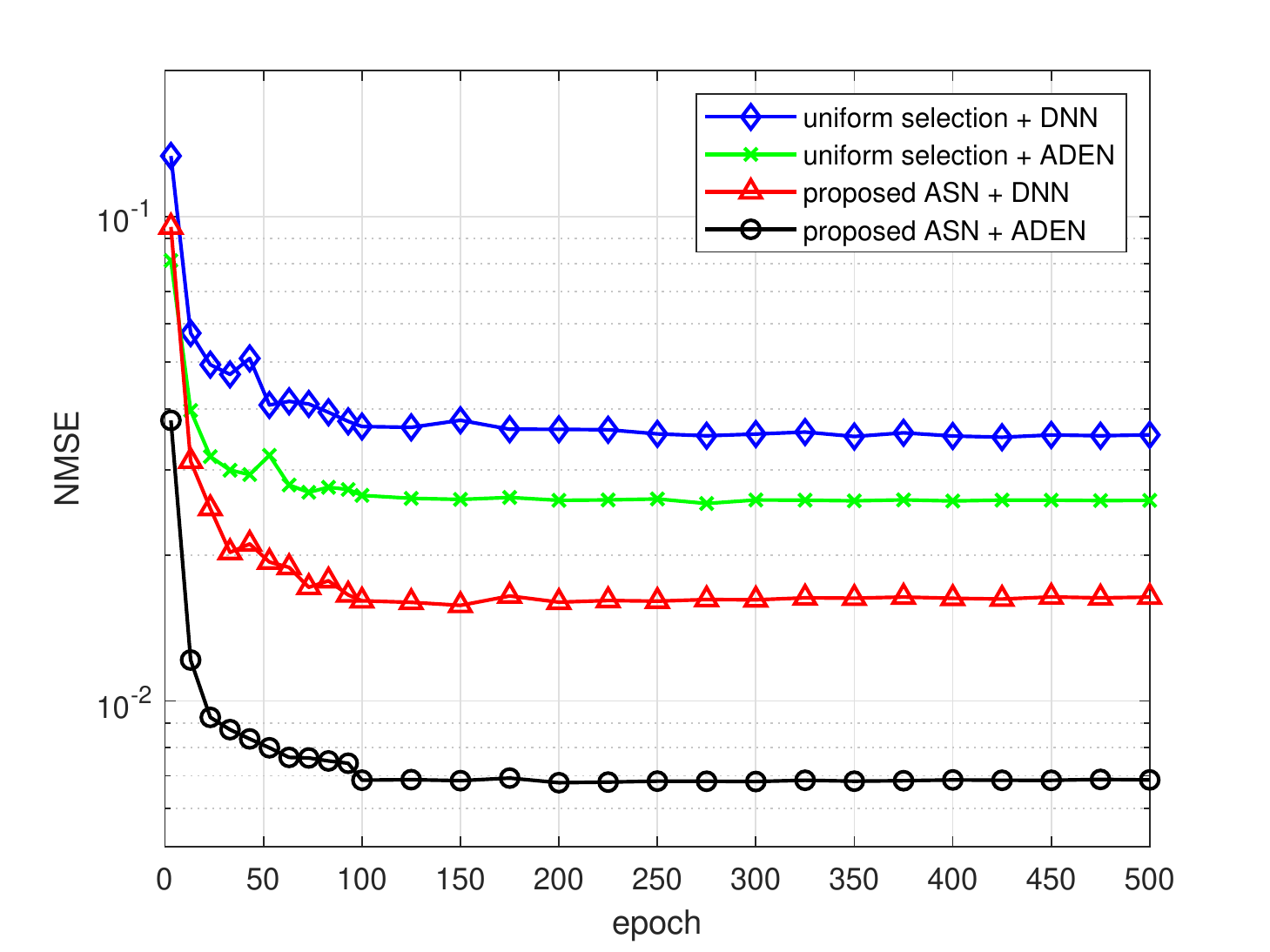}
\caption{The NMSE of CCM extrapolation  versus epoches with 8 antennas.}
\label{CCM-epoch}
\end{figure}
\begin{figure}[t]
\centering
\includegraphics[width=0.6\textwidth]{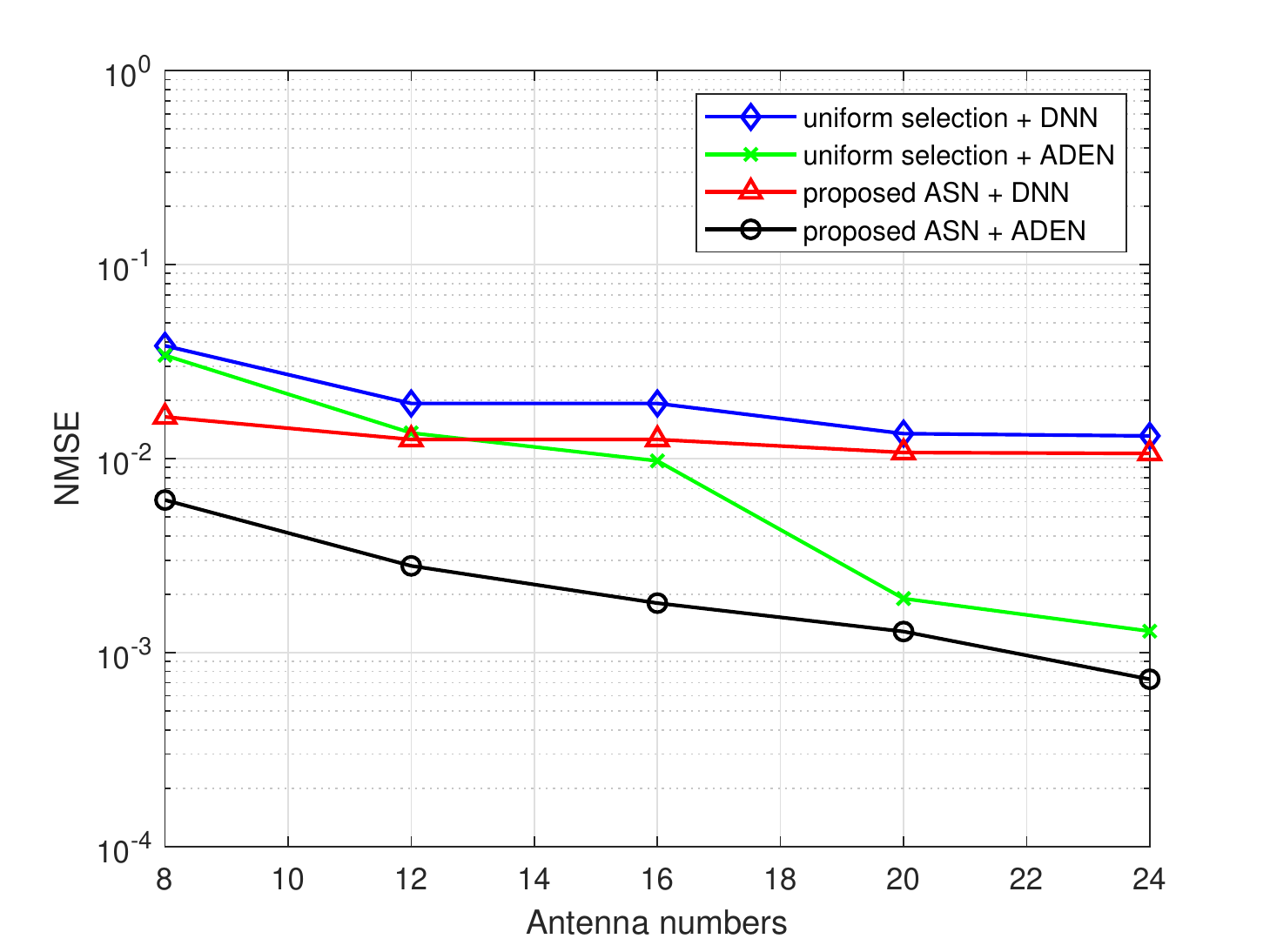}
\caption{The NMSE of CCM extrapolation versus antenna numbers.}
\label{CCM-antenna}
\end{figure}

For CCM extrapolation, we first use the CCM of 8 antennas to extrapolate the CCM of 64 antennas.
The antenna selection patterns for CCM extrapolation learned by `ASN + DNN' and `ASN + ADEN' are shown in Fig. \ref{ccm-pattern}, and
the NMSE of channel extrapolation for four different schemes are displayed in Fig. \ref{CCM-epoch}.
It is seen that the  NMSE of CCM extrapolation for `Uniform + DNN', `Uniform + ADEN', `ASN + DNN', and
`ASN + ADEN' are 0.036, 0.026, 0.016, 0.007 respectively.
We then show the NMSE of CCM extrapolation using different numbers of antennas in Fig. \ref{CCM-antenna}.
It is seen that the improvement brought by the antenna selection is quite significant when  $|\mathcal{B}|$  is small. Nevertheless, when $|\mathcal{B}|$ increases, the improvement brought by the antenna selection reduces while the improvement of extrapolation mostly comes from the designed ADEN.
Moreover, the proposed `ASN+ADEN' always achieves the best extrapolation accuracy with different antenna numbers.

\subsection{Sensitivity Analysis}
By configuring the neural networks with different initial weights, we will obtain different antenna selection patterns.
It is meaningful to study the variance of extrapolation error under different initialization conditions.
We repeat  training the `ASN + ADEN' for channel extrapolation with different antenna numbers several times, and compute the variances as depicted in Table~\ref{table3}.
Although the learned antenna selection pattern from each training is different, the extrapolation error does not fluctuate much, and the error variance  will be further reduced as the number of antennas increases.
\begin{table*}[t]
\renewcommand\arraystretch{1.2}
\caption{Channel Extrapolation Error with Different Initial Weights}\label{table3}
\centering
\begin{tabular}{m{2cm}<{\centering}|m{1cm}<{\centering}|m{1cm}<{\centering}|m{1cm}<{\centering}|m{1cm}<{\centering}|m{1cm}<{\centering}|m{1cm}<{\centering}|m{1cm}<{\centering}|m{1cm}<{\centering}|m{2cm}<{\centering}}
\hline
\hline
Antenna Number & \multicolumn{8}{c|}{Channel Extrapolation NMSE ($\times$e-03)} & Variance \\
\hline
8 & 6.936 & 7.220  &  8.426 & 4.845 & 7.062 & 6.714 & 8.920 & 6.670 & 1.313e-06\\
\hline
16 & 2.631 & 3.455 &  2.821 & 4.006 & 4.397 & 4.612 & 1.878 & 3.960 & 7.943e-07\\
\hline
24 & 1.813 & 1.494 &  2.675 & 1.621 & 2.315 & 1.707 & 2.528 & 2.269 & 1.755e-07\\
\hline
32 & 1.414 & 1.412 &  1.742 & 1.405 & 1.629 & 1.160 & 1.379 & 1.446 & 2.645e-08\\
\hline
\hline
\end{tabular}
\end{table*}

\section{Conclusions}\label{section6}
In this paper, we investigated the antenna domain channel extrapolation for massive MIMO system, where the channels of the whole antenna array can be predicted from that of a few antennas. We first designed the ASN to achieve the optimal antenna selection, where we proposed a constrained degradation method to approximate the derivative of the antenna selection vector such that the gradient can be back propagated when training the network.
We next design the ADEN to complete the channel extrapolation, where the ODE-inspired network structure is adopted to enhance the performance compared to the conventional DNN. The ASN and ADEN are jointly trained to find the optimal parameters. We then present three typical applications: channel   extrapolation, CCM extrapolation, and beam prediction. Simulations results show that the learned antenna selection is superior to the uniform selection, and the ADEN performs better than the tradition DNN.

\begin{appendices}
\section{Proof of Lemma 1}\label{appendix_proof}
\begin{proof}\label{proof1}
Since $\bm\nu^{'}$ is a rearrangement of $\bm\nu$, it still satisfies the equality constraints (\ref{constraint}).
Without loss of generality, we assume that there are $r$ ($0\textless r\leq N_{t}, r\in \mathbb{Z}$) non-zero elements in $\bm\nu^{'}$, i.e., $\nu_{k_{1}}\geq\nu_{k_{2}}\geq\cdots\nu_{k_{r}}\textgreater0$ and $\nu_{k_{r+1}}=\nu_{k_{r+2}}=\cdots\nu_{k_{N_{t}}}=0$.
Denote $\bm\nu_{r}^{'}=\left[\nu_{k_{1}},\nu_{k_{2}},\cdots,\nu_{k_{r}}\right]^{T}$.
Note that $\bm\nu_{r}^{'}$ still satisfies the equality constraints (\ref{constraint}).

For each $\nu_{k_{j}}$ in $\bm\nu_{r}^{'}$, we have $\nu_{k_{j}}\textgreater 0$, and hence there are  $\nu_{k_{j}}^{2},\nu_{k_{j}}^{3}\textgreater 0$.
According to Cauchy Schwarz inequality \cite{steele2004cauchy}, we obtain
\begin{equation}\label{cauch}
\left(\nu_{k_{1}}+\nu_{k_{2}}+\cdots+\nu_{k_{r}}\right)\left(\nu_{k_{1}}^{3}+\nu_{k_{2}}^{3}+\cdots+\nu_{k_{r}}^{3}\right)\geq\left[\left(\nu_{k_{1}}^{2}+\nu_{k_{2}}^{2}+\cdots+\nu_{k_{r}}^{2}\right)\right]^{2},
\end{equation}
where the the equality holds if and only if $\left[\nu_{k_{1}},\cdots,\nu_{k_{N_{t}}}\right]^{T}$ and $\left[\nu_{k_{1}}^{3},\cdots,\nu_{k_{r}}^{3}\right]^{T}$ are linearly dependent.

From (\ref{constraint}), we know
\begin{equation}\label{equal}
\left(\nu_{k_{1}}+\nu_{k_{2}}+\cdots+\nu_{k_{r}}\right)\left(\nu_{k_{1}}^{3}+\nu_{k_{2}}^{3}+\cdots+\nu_{k_{r}}^{3}\right)=\left[\left(\nu_{k_{1}}^{2}+\nu_{k_{2}}^{2}+\cdots+\nu_{k_{r}}^{2}\right)\right]^{2}=M_{t}^{2}.
\end{equation}

Hence, the equality (\ref{cauch}) holds, and we obtain
\begin{equation}\label{nu_equation}
\frac{\nu_{k_{1}}^{3}}{\nu_{k_{1}}}=\frac{\nu_{k_{2}}^{3}}{\nu_{k_{2}}}=\cdots=\frac{\nu_{k_{r}}^{3}}{\nu_{k_{r}}}=q.
\end{equation}

Solving (\ref{nu_equation}), we obtain  $\nu_{k_{1}}=\nu_{k_{2}}=\cdots=\nu_{k_{r}}=\sqrt{q}$.
Substitute $\nu_{k_{i}}=\sqrt{q} (i=1,\cdots,r)$ into (\ref{constraint}), we obtain the following equation
\begin{equation}\label{equ}
\left\{
\begin{aligned}
&r\cdot \sqrt{q}=M_{t}\\
&r\cdot (\sqrt{q})^{2}=M_{t}\\
\end{aligned}
\right..
\end{equation}

Solving equation (\ref{equ}), we obtain $q=1,r=M_{t}$, and there are $\nu_{k_{1}}=\nu_{k_{2}}=\cdots=\nu_{k_{M_{t}}}=1$ and $\nu_{k_{M_{t}+1}}=\nu_{k_{M_{t}+2}}=\cdots=\nu_{k_{N_{t}}}=0$.
Therefore the vector $\bm\nu$ is an $M_{t}$-hot vector under the constraints (\ref{constraint}).
\end{proof}
\section{Geometric Explanation of Lemma 1}\label{appendix_lemma}
\begin{figure}[t]
\centering
\includegraphics[width=0.58\textwidth]{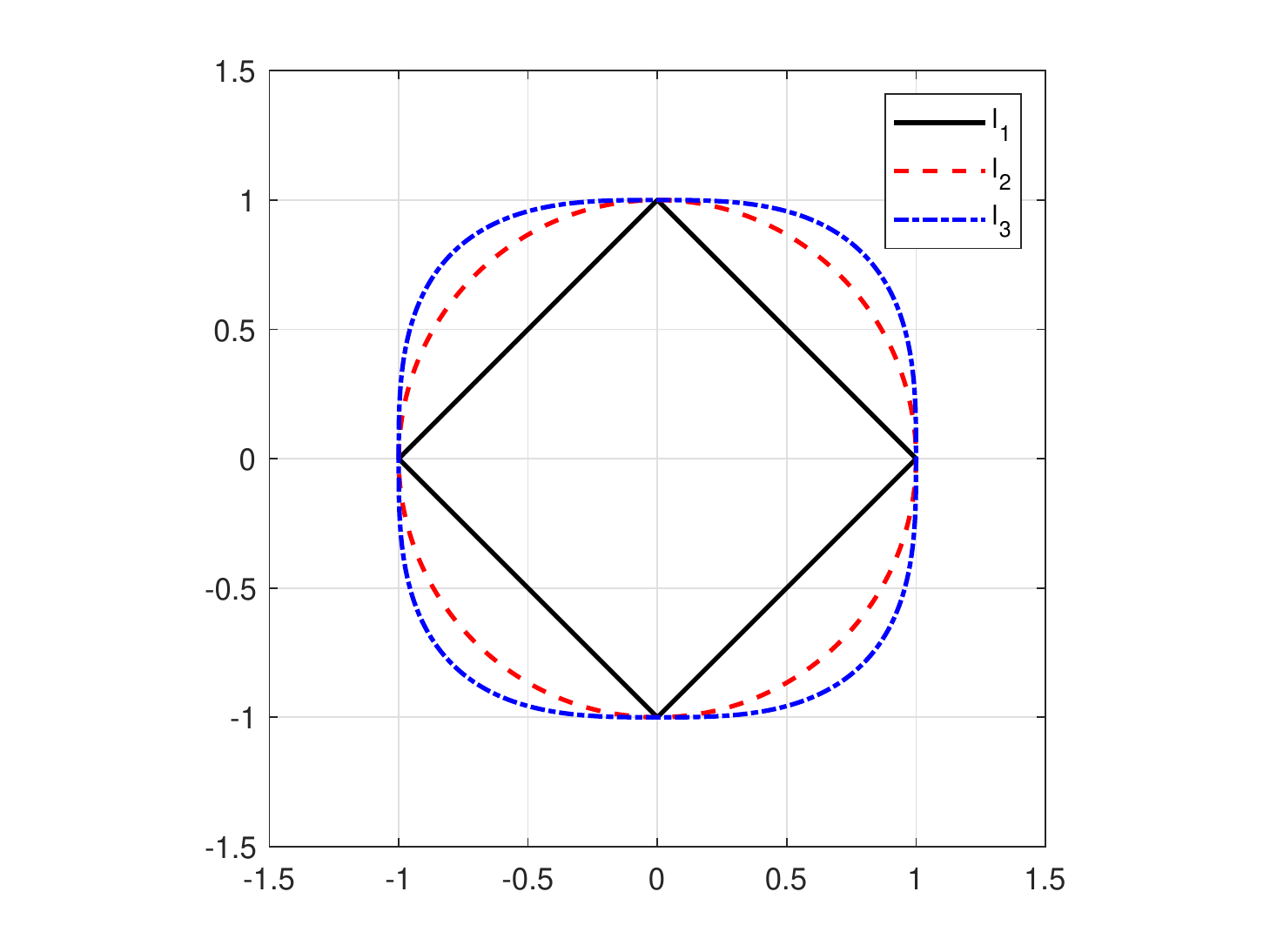}
\caption{Norm balls of 2-dimension vectors}
\label{norm_ball}
\end{figure}

\begin{figure}[t]
\centering
\includegraphics[width=0.56\textwidth]{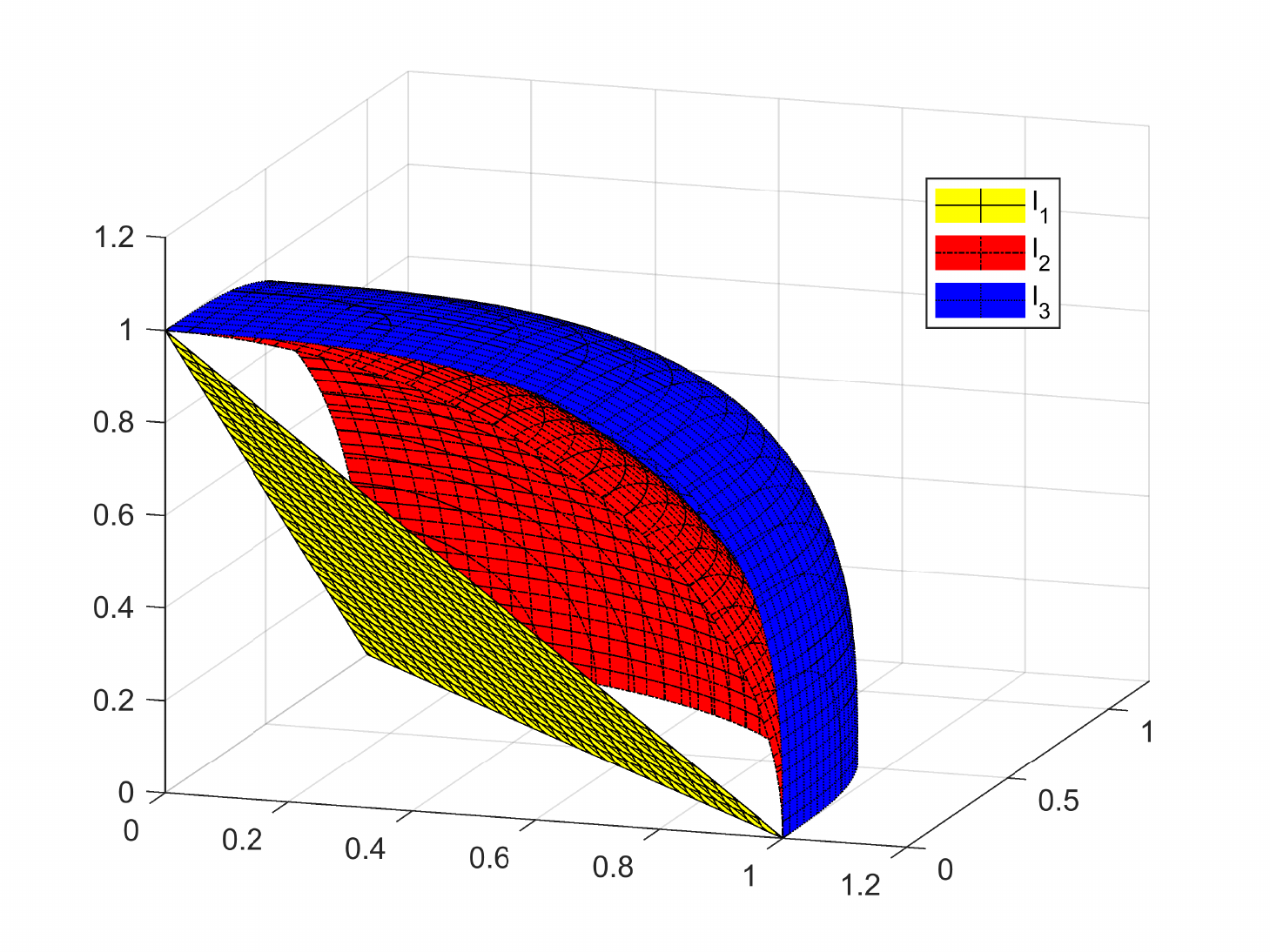}
\caption{Norm balls of 3-dimension vectors}
\label{norm_ball_3}
\end{figure}

We first display the $l_{1}$-, $l_{2}$-, and $l_{3}$-norm balls of two-dimension vectors   in Fig. \ref{norm_ball}.
It is seen that the intersections of these norm balls are all on x-axis and y-axis with  coordinates $(1,0)$, $(0,1)$, $(-1,0)$, $(0,-1)$.
If we restrict the horizontal and vertical coordinates to be non-negative numbers,  then there are only two intersections  $(1,0)$ and $(0,1)$ whose coordinates are exactly two one-hot vectors.
 We then display  the norm balls of three-dimension vectors  in Fig. \ref{norm_ball_3}, and   there are three intersections $(1,0,0)$, $(0,1,0)$ and $(0,0,1)$.
We see that the intersections of the $l_{1}$-norm, $l_{2}$-norm and $l_{3}$-norm balls are three one-hot vectors.
\begin{figure}[t]
\centering
\includegraphics[width=0.56\textwidth]{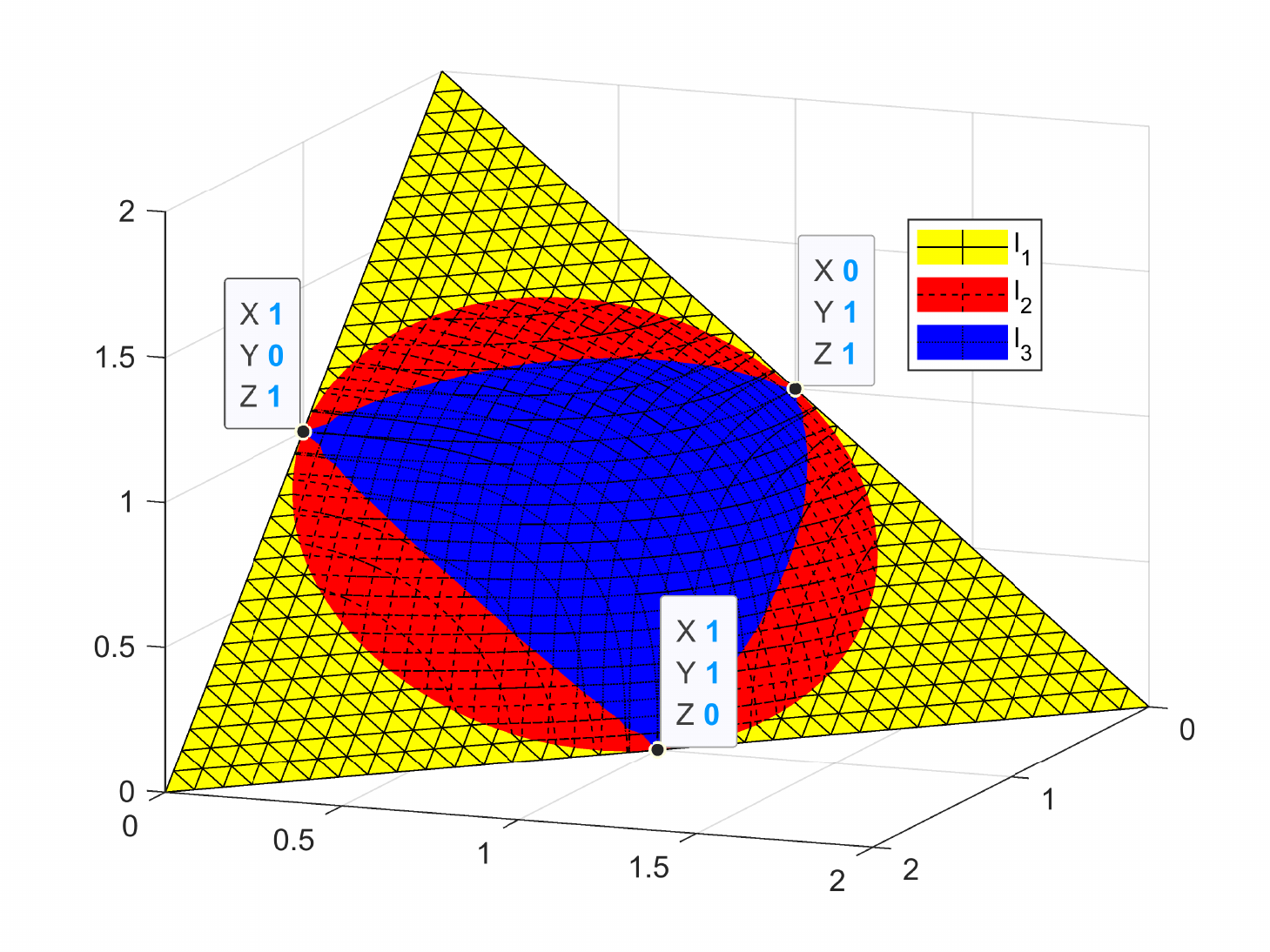}
\caption{Norm balls of 3-dimension vectors: $\|\bm x\|_{p}=\sqrt[p]{2}$}
\label{norm_ball_3_2}
\end{figure}

Next, we show how to yield a K-hot vector by deforming the norm balls.
Taking three-dimension vectors as an example, we wish to find several graphics such that the coordinates of their intersections are two-hot vectors (1,1,0), (0,1,1), (1,0,1), respectively.
Since the $l_{p}$-norm of a two-hot vector is $\sqrt[p]{2}$,  we could construct the following three graphics
\begin{equation}\label{two-hot}
\begin{aligned}
\begin{aligned}
\|\bm x\|_{1}^{1}=2,\quad
\|\bm x\|_{2}^{2}=2,\quad
\|\bm x\|_{3}^{3}=2,
\end{aligned}
\end{aligned}
\end{equation}
as shown  in Fig. \ref{norm_ball_3_2}. It seen from Fig. \ref{norm_ball_3_2} that, the coordinates of three intersections from the three graphics are exactly the two-hot vectors we need.

For more general case, by limiting the $l_{1}$-norm, $l_{2}$-norm, and $l_{3}$-norm of the vectors to $\sqrt[p]{K}$, we can get K different K-hot vectors. The strict proof can be found in Section \ref{section3}.
\section{Derivation of ODE}\label{appendix_ode}
We here derive the mathematical model of the fine extrapolation subnetwork.
The target is to obtain the fine $\hat{\mathbf{u}}_{\mathcal{A}}^{f}$ from the coarse $\hat{\mathbf{u}}_{\mathcal{A}}^{c}$.
A popular way to improve the prediction accuracy is to increase the depth of DNN.
However, simply increasing the depth of DNN may bring various issues like overfitting, vanishing gradient, etc.
In this sense, many neural network structures \cite{he2016deep,he2016identity,he2019ode} based on skip-connections \cite{mao2016image} were proposed to help increase the layers of the network and have achieved advanced performance.
The skip connection can be formulated as
\begin{equation}\label{skip_connection}
\mathbf{u}_{f}(n+1)=\mathbf{u}_{f}(n)+\bm\psi\left[\mathbf{u}_{f}(n),\bm\omega(n)\right],
\end{equation}
where $n+1$, $n$ are the layer indices  in DNN (also can be seen as the time mark) and $\mathbf{u}_{f}(n+1)$, $\mathbf{u}_{f}(n)$ are two neuron layers.
Equation (\ref{skip_connection}) is also known  as the  discretized Euler equation \cite{lu2018beyond}.
Replacing discrete variable $n$ with continuous variable $t$, when the time interval becomes small ($t\rightarrow0$) (or the number of layers between the connected layers becomes large), equation (\ref{skip_connection}) can be written as
\begin{equation}\label{x_ode}
\frac{d\mathbf{u}_{f}(t)}{dt}=\bm\psi\left[\mathbf{u}_{f}(t),\bm\omega(t)\right].
\end{equation}
For the  fine extrapolation subnetwork, the input is $\mathbf{u}_{f}(0)=\hat{\mathbf{u}}_{\mathcal{A}}^{c}$.
Then fine extrapolation can be formulated as an ordinary differential equation (ODE) initial value problem
\begin{equation}\label{ode_initial_appendix}
\left\{
\begin{aligned}
&\frac{d\mathbf{u}_{f}(t)}{dt}=\bm\psi\left[\mathbf{u}_{f}(t),\bm\omega(t)\right]\\
&\mathbf{u}_{f}(0)=\mathbf{u}_{\mathcal{B}}\\
\end{aligned}
\right..
\end{equation}
\end{appendices}

\linespread{1.4}

\bibliographystyle{IEEEtran}

\end{document}